\documentclass[aps,onecolumn,nofootinbib,superscriptaddress,tightenlines]{revtex4}

\usepackage{amsmath, amsthm, amssymb, slashed}
\usepackage{amsfonts}
\usepackage[latin1]{inputenc}
\usepackage{MnSymbol}
\usepackage{graphicx}
\usepackage{dcolumn}
\usepackage{tabularx}
\usepackage{leftidx}
\usepackage{fancyheadings}
\usepackage{boxedminipage}
\usepackage{float}
\usepackage{booktabs}
\usepackage{dsfont} 
\usepackage{tensor}
\usepackage{fancyvrb}
\usepackage{verbatim}
\usepackage{enumitem}
\usepackage{mathrsfs}
\usepackage[hidelinks]{hyperref}
\usepackage{wrapfig}

\def\tilde{\widetilde}
\def\spb#1{\rlap{\lower1.5ex \hbox{$\leftarrow$}}{#1}}
\def\sdpb#1{\rlap{\lower1.5ex \hbox{$\Leftarrow$}}{#1}}

\DeclareMathAlphabet{\mathscrbf}{OMS}{mdugm}{b}{n}

\parskip=\baselineskip

\begin{document}

\title{Kerr Isolated Horizons in Ashtekar and Ashtekar--Barbero Connection Variables}

\vskip.9in

\author{Christian R\"oken}

\vskip.5in

\affiliation{Universit\"at Regensburg, Fakult\"at f\"ur Mathematik, 93040 Regensburg, Germany \footnotetext{e-mail: christian.roeken@mathematik.uni-regensburg.de}}
\vskip.2in 
\affiliation{Centre de Physique Th\'{e}orique de Luminy, Case $907$, $13288$ Marseille, France}

\vskip.9in

\date{July 2017}

\begin{abstract}
\vskip.2in 
\noindent \noindent \textbf{\footnotesize ABSTRACT.} \, The Ashtekar and Ashtekar--Barbero connection variable formulations of Kerr isolated horizons are derived. Using a regular Kinnersley tetrad in horizon-penetrating Kruskal--Szekeres-like coordinates, the spin coefficients of Kerr geometry are determined by solving the first Maurer--Cartan equation of structure. Isolated horizon conditions are imposed on the tetrad and the spin coefficients. A transformation into an orthonormal tetrad frame that is fixed in the time gauge is applied and explicit calculations of the spin connection, the Ashtekar and Ashtekar--Barbero connections, and the corresponding curvatures on the horizon $2$-spheres are performed. Since the resulting Ashtekar--Barbero curvature does not comply with the simple form of the horizon boundary condition of Schwarzschild isolated horizons, i.e., on the horizon $2$-spheres, the Ashtekar--Barbero curvature is not proportional to the Plebanski $2$-form, which is required for an SU(2) Chern-Simons treatment of the gauge degrees of freedom in the horizon boundary in the context of loop quantum gravity, a general method to construct a new connection whose curvature satisfies such a relation for Kerr isolated horizons is introduced. For the purpose of illustration, this method is employed in the framework of slowly rotating Kerr isolated horizons.
\end{abstract}

\maketitle

\tableofcontents

\section{Introduction} \label{SecI}

\noindent In general relativity, a charged, rotating black hole in equilibrium is usually described in terms of the future event horizon of the Kerr--Newman solution of the vacuum Einstein field equations despite the fact that the canonical definition of a future event horizon is a non-local concept, i.e., it refers to future null infinity and requires the existence of a global time-translational Killing vector field for static spacetimes and an asymptotic time-translational Killing vector field at space-like infinity for stationary spacetimes. This concept is overly restrictive because it suffices to consider a horizon structure with a local time-translational Killing vector field, that is, with a time-independent intrinsic geometry and the possibility for a dynamical bulk geometry. Moreover, since the information on the entire spacetime history is needed when one refers to an event horizon, it is in general not well-suited for describing black hole evolution. A more appropriate notion, however, is given by the so-called quasi-local isolated horizons, which account for equilibrium states of black holes and cover all essential local features of event horizons (for reviews on the subject see, e.g., \cite{AK, DPP}). Isolated horizons employ only local time-translational Killing vector fields and require neither asymptotic structures nor foliations of spacetime. Consequently, they allow for the occurrence and for the dynamics of matter and radiation distributions in the exterior spacetime. In the following, we recall the definition of isolated horizons. 

An isolated horizon $\Delta$ is a $3$-dimensional submanifold of a $4$-dimensional globally hyperbolic, asymptotically flat spacetime $(\mathfrak{M}, \boldsymbol{g})$ that is equipped with an equivalence class 
\begin{equation}\label{ECNN} 
\left[\boldsymbol{X}\right]_{\sim} := \{\boldsymbol{X} \in T\mathfrak{M} \, | \, \boldsymbol{X} = c \, \boldsymbol{X}', \, c \in \mathbb{R}_{> 0} \} 
\end{equation}
of normal, future-directed null vector fields and satisfies the conditions: 
\begin{enumerate}
\item[(i)] $\Delta$ is a null $3$-surface with topology $S^2 \times \mathbb{R}$, which admits a suitable foliation by a family of $2$-spheres.
\item[(ii)] The outgoing expansion rate $\theta_{(\boldsymbol{X})} = q_{\mu \nu} \nabla^{\mu} X^{\nu}$ of the vector fields $\boldsymbol{X} \in \left[\boldsymbol{X}\right]_{\sim}$, where $\boldsymbol{q}$ is the induced metric on the horizon and $\boldsymbol{\nabla}$ is the covariant derivative on $T\mathfrak{M}$, vanishes on $\Delta$. 
\item[(iii)] All field equations hold on $\Delta$. 
\item[(iv)] The stress-energy tensor $\boldsymbol{T}$ on $\Delta$ satisfies the dominant energy condition. 
\item[(v)] The induced derivative operator $\boldsymbol{\mathfrak{D}}$ on $\Delta$, which is defined by the horizon pullback of $\boldsymbol{\nabla}$ and the vector fields $\boldsymbol{Y}, \boldsymbol{Z} \in T\Delta$ via
$Y^{\mu} \mathfrak{D}_{\mu} Z^{\nu} := Y^{\mu} \spb{\hspace{0.03cm}\nabla}_{\mu} Z^{\nu}$, satisfies
\begin{itemize}
\item[--] $\bigl[\mathfrak{L}_{\boldsymbol{X}} , \boldsymbol{\mathfrak{D}}\bigr] \boldsymbol{X} = 0 \,\,\,\,\, \forall \,\,\,\,\, \boldsymbol{X} \in \left[\boldsymbol{X}\right]_{\sim} \,\,\, (\textnormal{weakly isolated horizon condition})$
\item[--] $\bigl[\mathfrak{L}_{\boldsymbol{X}} , \boldsymbol{\mathfrak{D}}\bigr] \boldsymbol{Y} = 0 \,\,\,\,\,\, \forall \,\,\,\,\, \boldsymbol{Y} \in T\Delta \,\,\,\,\,\,\,\,\, (\textnormal{isolated horizon condition})$\,,
\end{itemize}   
where the bracket $\left[\cdot \, , \cdot\right]$ denotes the commutator and $\mathfrak{L}_{\boldsymbol{X}}$ the Lie derivative along the null vector field $\boldsymbol{X}$. 
\end{enumerate}
Using the Raychaudhuri equation and the dominant energy condition, one can prove that 
\begin{equation} \label{IHMETRICCOND}
\mathfrak{L}_{\boldsymbol{X}} q_{\mu \nu} = 0 \, . 
\end{equation} 
This equation shows that the null normal $\boldsymbol{X}$ is a local time-translational Killing vector field on the horizon. Furthermore, it expresses the time independence of the induced metric. The time independence of the induced derivative operator is directly given by the isolated horizon condition in (v). Thus, the intrinsic isolated horizon geometry can be captured by the pair $(\boldsymbol{q}, \boldsymbol{\mathfrak{D}})$. A Carter--Penrose diagram of a spacetime containing an isolated horizon is shown in Figure 1. Isolated horizons can be classified into three different types, namely spherically symmetric horizons, axisymmetric horizons, and general, distorted horizons (for more details, in particular on their symmetry groups, see \cite{ABL0, LewPaw}). The basic spherically symmetric type has already been analyzed extensively in both the classical and quantum regimes (see, e.g., \cite{ABCK, ACK, ABK, AFK, ABL, ENPP, HS}), whereas studies concerning the more intricate axisymmetric and distorted types \cite{AEV, LewPaw0, PANPER} are up to now still incomplete. In the present article, we provide a well-behaved set of connection variables for Kerr isolated horizons, i.e., we explicitly compute expressions for the Ashtekar and Ashtekar--Barbero connections and curvatures with a focus on regularity at the event horizon region. Furthermore, we show that the horizon $2$-sphere pullback of the Ashtekar--Barbero curvature is not proportional to that of the Plebanski $2$-form, which is in contrast to the associated horizon boundary condition of the Schwarzschild isolated horizon case. As such a condition relates the degrees of freedom of the bulk geometry to the isolated horizon surface degrees of freedom, making it the key constraint in the classical symplectic theory, and because it is essential for an SU($2$) Chern--Simons treatment of the gauge degrees of freedom in the horizon boundary in the context of loop quantum gravity, it is of great interest to have it satisfied also for Kerr isolated horizons. Hence, we present a non-standard method for the construction of a new connection whose curvature fulfills this particular form of the horizon boundary condition for Kerr isolated horizons.

%
%
\begin{wrapfigure}{r}{0.44\textwidth}
	\begin{center}
		\includegraphics[width=0.26\textwidth]{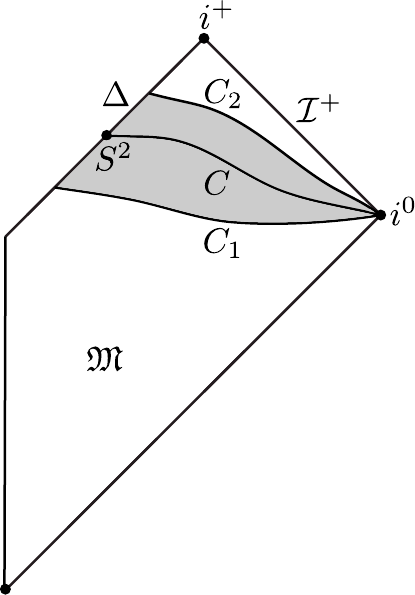}
		\caption{Carter--Penrose diagram of a spacetime $(\mathfrak{M}, \boldsymbol{g})$ containing an isolated horizon $\Delta$. The shaded region of $\mathfrak{M}$ that is bounded by $\Delta$ and the partial Cauchy surfaces $C_1$ and $C_2$, which start at space-like infinity $i^0$ and intersect the horizon, accounts for a black hole in equilibrium. The intersection $S^2 = \Delta \bigcap C$ of the isolated horizon with the intermediate Cauchy surface $C$ is of spherical topology.} 
	\end{center}
\end{wrapfigure}
%
%

In more detail, we are concerned with the rotating Kerr-type of the isolated horizon models, which represents one specific symmetric rotator amongst the many rotating isolated horizon geometries. Nonetheless, it is of particular interest since, in the astrophysical context, it accounts for supermassive spinning black holes and the associated accretion disks in active galactic nuclei in good approximation. Besides, recent observations in astrophysics \cite{MC} indicate that most of the black holes actually appear to rotate quite rapidly around an internal symmetry axis and, thus, probably constitute the most common form existing in nature. The general idea behind Kerr isolated horizons is the following. Every Killing horizon that has the topology $S^2 \times \mathbb{R}$ is an isolated horizon \cite{AK}. This implies that in particular the event horizon of Kerr geometry is an isolated horizon. However, working with such a horizon leads to the aforementioned problems, namely that one usually requires an asymptotic time-translational Killing vector field at space-like infinity and refers to future null infinity. Thus, we relax the first condition by using only a local time-translational Killing vector field at the horizon, given in terms of the equivalence class of local null normal vector fields (\ref{ECNN}). The latter condition is dropped completely, as we only consider quantities that are intrinsic to the horizon. The starting point of our analysis is Kerr geometry represented by a regular Kinnersley tetrad in horizon-penetrating Kruskal--Szekeres-like coordinates, which covers the exterior as well as the interior black hole region, up to the Cauchy horizon. Solving the first Maurer--Cartan equation of structure in this frame, we determine the spin coefficients. Both the tetrad and the spin coefficients are then adapted to the isolated horizon conditions using local Lorentz transformations. With the resulting quantities, we compute the spin connection on the horizon $2$-spheres. This spin connection is subsequently transformed into an orthonormal frame that is fixed in the time gauge. Next, we explicitly calculate the Ashtekar and Ashtekar--Barbero connections on the horizon $2$-spheres, which are defined by linear combinations of the components of the spin connection, and the corresponding curvatures. It turns out, as already noted above, that the resulting Ashtekar--Barbero curvature does not comply with the simple form of the horizon boundary condition of the Schwarzschild isolated horizon case, that is, it is not proportional to the Plebanski $2$-form projected onto the horizon $2$-spheres. Therefore, we introduce a method for the construction of a new connection whose curvature satisfies a horizon boundary condition of the desired form. For illustrative purposes, this method is used in the framework of slowly rotating Kerr isolated horizons, i.e., in the first-order formalism for small angular momenta.

\section{Preliminaries}

\noindent Kerr geometry is described by a connected, orientable and time-orientable, smooth, asymptotically flat Lorentzian $4$-manifold ($\mathfrak{M}, \boldsymbol{g}$) with topology $S^2 \times \mathbb{R}^2$, in which the metric $\boldsymbol{g}$ is stationary and axisymmetric. The tangent and cotangent bundles $T\mathfrak{M}$ and $T^{\star}\mathfrak{M}$ are endowed with the respective bases $(\boldsymbol{e}_{\mu})$ and $(\boldsymbol{e}^{\mu})$, $\mu \in \{0, 1, 2, 3\}$, and with the Levi--Civita connection $\boldsymbol{\omega}$. Choosing coordinates $(x^{\mu}) = (t = x^0, x^1, x^2, \varphi = x^3)$, where $t$ is the coordinate time and $\varphi$ the azimuthal angle about the axis of symmetry, for which the Killing vectors are given by $\partial_t$ and $\partial_{\varphi}$ and all metric coefficients $g_{\mu \nu} = g_{\mu \nu}(x^1, x^2)$ are independent of $t$ and $\varphi$, the metric becomes
\begin{equation*}
\begin{split}
\boldsymbol{g} & = g_{t t} \, \textnormal{d}t \otimes \textnormal{d}t + g_{x^1 x^1} \,  \textnormal{d}x^1 \otimes \textnormal{d}x^1 + g_{x^1 x^2} \, \bigl(\textnormal{d}x^1 \otimes \textnormal{d}x^2 + \textnormal{d}x^2 \otimes \textnormal{d}x^1\bigr) + g_{x^2 x^2} \, \textnormal{d}x^2 \otimes \textnormal{d}x^2 \\ 
& \hspace{0.3cm} + g_{t \varphi} \, \bigl(\textnormal{d}t \otimes \textnormal{d}\varphi + \textnormal{d}\varphi \otimes \textnormal{d}t\bigr) + g_{\varphi \varphi} \,  \textnormal{d}\varphi \otimes \textnormal{d}\varphi \, .
\end{split}
\end{equation*}
Note that in order to obtain this particular representation, one additionally requires the metric to be invariant under simultaneous discrete time and azimuthal angle isometries $t \mapsto - t$ and $\varphi \mapsto - \varphi$. In the present study, we employ horizon-penetrating Kruskal--Szekeres-like coordinates (Appendix \ref{AA2}) and express Kerr geometry in terms of a regular Kinnersley tetrad (Appendix \ref{AA3}), i.e., a local Newman--Penrose null frame that is, on the one hand, adapted to the class of principal null geodesics and, on the other hand, regular at the event horizon. (For more details on the Newman--Penrose formalism see Appendix \ref{AA}.) Since Kerr geometry is algebraically special and of Petrov type D, using the Kinnersley frame has the computational advantage that one has four vanishing spin coefficients ($\kappa = \sigma = \lambda = \nu = 0$) and only one non-vanishing Weyl scalar ($\Psi_2 \not= 0$). Moreover, one of the elements of this frame coincides with the isolated horizon generator (cf.\ Appendix \ref{AD}).

\section{Spin Connection for Kerr Isolated Horizons}

\noindent We begin by imposing the isolated horizon conditions on the regular Kinnersley tetrad (\ref{NPBV}) (Appendix \ref{AD}). Then, using the corresponding spin coefficients (\ref{DPSCOEF}), we compute the pullback of the spin connection to the isolated horizon $2$-spheres and apply a transformation into an orthonormal frame that is fixed in the time gauge (Appendix \ref{A7}). The resulting spin connection reads
\begin{equation}\label{pbsc}
\begin{split}
\sdpb{\hspace{0.07cm}{\omega}_{\hspace{0.04cm} (0) (1)}} & = \frac{a \sin{(\theta)}}{\rho_+^2} \, \Biggl(a \cos{(\theta)} \, \textnormal{d}\theta +\rho_{0 +}^2 \sin{(\theta)} \left[\alpha_+ + \frac{r_+}{\rho_+^2}\right] \textnormal{d}\tilde{\varphi}_+\Biggr) \\ \\
\sdpb{\hspace{0.07cm}{\omega}_{\hspace{0.04cm} (0) (2)}} & = \sdpb{\hspace{0.07cm}{\omega}_{\hspace{0.04cm} (1) (2)}} = - \frac{1}{\sqrt{2} \, \rho_+^4} \, \Biggl(\left[r_+^2 \rho_{0 +}^2 - \frac{a^4 \sin^2{(2 \theta)}}{4}\right] \textnormal{d}\theta + \frac{a \sin{(2 \theta)} \rho_{0 +}^2}{2} \left[r_+ - a^2 \sin^2{(\theta)} \left(\alpha_+ + \frac{2 r_+}{\rho_+^2}\right)\right] \textnormal{d}\tilde\varphi_+\Biggr)\\ \\
\sdpb{\hspace{0.07cm}{\omega}_{\hspace{0.04cm} (0) (3)}} & = \sdpb{\hspace{0.07cm}{\omega}_{\hspace{0.04cm} (1) (3)}} = \frac{1}{\sqrt{2} \, \rho_+^4} \, \Biggl(r_+ a \cos{(\theta)} \left[\rho_{0 +}^2 + a^2 \sin^2{(\theta)}\right] \textnormal{d}\theta \\ \\
& \hspace{3.3cm} - \rho_{0 +}^2 \sin{(\theta)} \left[r_+^2 - a^2 \sin^2{(\theta)} \left(r_+ \alpha_+ + \frac{r_+^2 - a^2 \cos^2{(\theta)}}{\rho_+^2}\right)\right] \textnormal{d}\tilde\varphi_+\Biggr)\\ \\
\sdpb{\hspace{0.07cm}{\omega}_{\hspace{0.04cm} (2) (3)}} & = \frac{1}{\rho_+^2} \, \biggl(- r_+ a \sin{(\theta)} \, \textnormal{d}\theta + \frac{\rho_{0 +}^4 \cos{(\theta)} }{\rho_+^2} \, \textnormal{d}\tilde\varphi_+\biggr) \, ,
\end{split}
\end{equation}
where $\theta \in [0, \pi]$ and $\tilde\varphi_+ \in [0, 2 \pi)$ are angular coordinates covering the horizon $2$-spheres, $a M$ is the angular momentum and $M$ the mass of the Kerr black hole, $r_{\pm} := M \pm \sqrt{M^2 - a^2}$ denote the event and the Cauchy horizon, respectively, $\rho_+^2 = \rho_+^2(\theta) :=  r_+^2 + a^2 \cos^2{(\theta})$, $\rho_{0 +}^2 := r_+^2 + a^2$, and $\alpha_+ := (r_+ - r_-)/(2 \rho^2_{0 +})$.
These quantities are the basis for the subsequent analysis of the Ashtekar and Ashtekar--Barbero connection variable formulations of Kerr isolated horizons.

\section{Ashtekar and Ashtekar--Barbero Connections and Curvatures}

\noindent In this section, we explicitly calculate the Ashtekar and Ashtekar--Barbero connections and curvatures on the horizon $2$-spheres (for a brief summary of the Ashtekar and Ashtekar--Barbero formulations see Appendix \ref{AE}). The self-dual Ashtekar connection $\boldsymbol{A}_+$ and the associated curvature $\boldsymbol{F}_+$ are defined as
\begin{equation}\label{AC2}
A^{(i)}_+ := \frac{1}{2} \, \epsilon\indices{^{(i)}_{(j) (k)}} \, \omega^{(j) (k)} + \textnormal{i} \, \omega^{(0) (i)} 
\end{equation}
and
\begin{equation}\label{CSDC}
F^{(i)}_+ := F^{(i)}(\boldsymbol{A}_+) = \textnormal{d}A^{(i)}_+ + \frac{1}{2} \, \epsilon\indices{^{(i)}_{(j) (k)}} \, A^{(j)}_+ \wedge A^{(k)}_+ 
\end{equation}
with $i, j, k \in \{1, 2, 3\}$. Substituting the spin connection $1$-forms (\ref{pbsc}) into (\ref{AC2}) and employing the convention $\epsilon^{(1) (2) (3)} = 1$, the Ashtekar connection on the horizon $2$-spheres becomes
\begin{equation}\label{ACUR}
\begin{split}
\sdpb{\hspace{0.04cm} A^{(1)}_{\hspace{0.03cm} +}} & = - \frac{a \sin{(\theta)}}{\overline{\varsigma}_+} \, \textnormal{d}\theta + \frac{\rho_{0 +}^2}{\rho_+^2} \Biggl[\cos{(\theta) - \textnormal{i} a \sin^2{(\theta)} \,  \biggl(\alpha_+ + \frac{1}{\overline{\varsigma}_+}\biggr)}\Biggr] \,  \textnormal{d}\tilde\varphi_+ \\ \\
\sdpb{\hspace{0.04cm} A^{(2)}_{\hspace{0.03cm} +}} & = - \textnormal{i} \sdpb{\hspace{0.04cm} A^{(3)}_{\hspace{0.03cm} +}} = \frac{1}{\sqrt{2} \, \overline{\varsigma}_+} \Biggl[\biggl(r_+ + \frac{a^2 \sin^2{(\theta)}}{\overline{\varsigma}_+} \biggr) \, \textnormal{i} \, \textnormal{d}\theta + \frac{\rho_{0 +}^2 \sin{(\theta)}}{\rho_+^2} \, \biggl(r_+ - a^2 \sin^2{(\theta)} \, \biggl[\alpha_+ + \frac{1}{\overline{\varsigma}_+} \biggr]\biggr) \, \textnormal{d}\tilde{\varphi}_+\Biggr] \, ,
\end{split}
\end{equation}
where $\overline{\varsigma}_+ = \overline{\varsigma}_+(\theta) := r_+ - \textnormal{i} a \cos{(\theta)}$. Applying (\ref{ACUR}) in (\ref{CSDC}), we immediately obtain the corresponding Ashtekar curvature 
\begin{equation}\label{TGC}
\begin{split}
\sdpb{\hspace{0.04cm} F^{(1)}_{\hspace{0.04cm} +}} & =  2 \, \rho_{0 +}^2 \sin{(\theta)} \, \Psi_{2 \, | \, r = r_+} \, \textnormal{d}\theta \wedge \textnormal{d}\tilde\varphi_+ \\ \\
\sdpb{\hspace{0.04cm} F^{(2)}_{\hspace{0.04cm} +}} & = - \textnormal{i} \sdpb{\hspace{0.04cm} F^{(3)}_{\hspace{0.04cm} +}} = - \frac{3 \textnormal{i} a \sin^2{(\theta)} \, \rho_{0 +}^2}{\sqrt{2} \, \overline{\varsigma}_+} \, \Psi_{2 \, | \, r = r_+} \, \textnormal{d}\theta \wedge \textnormal{d}\tilde\varphi_+ 
\end{split}
\end{equation}
with
\begin{equation*}
\Psi_{2 \, | \, r = r_+} = - \frac{M}{\rho_+^6} \, \Bigl(r_+ \bigl[r_+^2 - 3 a^2 \cos^2{(\theta)}\bigr] + \textnormal{i} a \cos{(\theta)} \, \bigl[3 r_+^2 - a^2 \cos^2{(\theta)}\bigr]\Bigr)  
\end{equation*}
being the second Weyl scalar evaluated at the event horizon $r = r_+$. In terms of the self-dual projection of the Plebanski $2$-form $\boldsymbol{\Sigma} := \boldsymbol{e} \wedge \boldsymbol{e}$, which in the time gauge is given by
\begin{equation*}
\Sigma^{(i)} = \frac{1}{2} \, \epsilon\indices{^{(i)}_{(j) (k)}} \, \boldsymbol{e}^{(j)} \wedge \boldsymbol{e}^{(k)} \, , 
\end{equation*}
the Ashtekar curvature (\ref{TGC}) yields
\begin{equation*}
\begin{split}
\sdpb{\hspace{0.04cm} F^{(1)}_{\hspace{0.04cm} +}} & = 2 \, \Psi_{2 \, | \, r = r_+} \, \sdpb{\hspace{0.04cm} \Sigma^{(1)}} \\ \\
\sdpb{\hspace{0.04cm} F^{(2)}_{\hspace{0.04cm} +}} & = - \textnormal{i} \sdpb{\hspace{0.04cm} F^{(3)}_{\hspace{0.04cm} +}} = 
- \frac{3 \textnormal{i} a \sin{(\theta)}}{\sqrt{2} \, \overline{\varsigma}_+} \, \Psi_{2 \, | \, r = r_+} \, \sdpb{\hspace{0.04cm} \Sigma^{(1)}} \, ,
\end{split}
\end{equation*}
where
\begin{equation*}
\sdpb{\hspace{0.04cm} \Sigma^{(1)}} = \rho_{0 +}^2 \sin{(\theta)} \, \textnormal{d}\theta \wedge \textnormal{d}\tilde\varphi_+ \,\,\,\,\,\, \textnormal{and} \,\,\,\,\,\, \sdpb{\hspace{0.04cm} \Sigma^{(2)}} = \sdpb{\hspace{0.04cm} \Sigma^{(3)}} = 0 \, . 
\end{equation*}
Next, computing the horizon $2$-sphere pullback of the Ashtekar--Barbero connection 
\begin{equation*}
A^{(i)}_{\gamma} := \frac{1}{2} \, \epsilon\indices{^{(i)}_{(j) (k)}} \, \omega^{(j) (k)} + \gamma \, \omega^{(0) (i)} \, , \,\,\,\,\,\, \gamma \in \mathbb{C} \backslash \{0\} \, ,
\end{equation*}
we find
\begin{equation}\label{ABCO}
\begin{split}
\sdpb{\hspace{0.04cm} A^{(1)}_{\hspace{0.03cm} \gamma}} & = - \frac{a \sin{(\theta)}}{\rho_+^2} \, \bigl[r_+ + \gamma a \cos{(\theta)}\bigr] \,  \textnormal{d}\theta + \frac{\rho_{0 +}^2}{\rho_+^2} \, \Biggl(\frac{\rho_{0 +}^2}{\rho_+^2} \, \cos{(\theta)} - \gamma a \sin^2{(\theta)} \, \biggl[\alpha_+ + \frac{r_+}{\rho_+^2}\biggr]\Biggr) \, \textnormal{d}\tilde{\varphi}_+ \\ \\
\sdpb{\hspace{0.04cm} A^{(2)}_{\hspace{0.03cm} \gamma}} & = \frac{1}{\sqrt{2} \, \rho_+^4} \, \Biggl[\Bigl(r_+ \rho_{0 +}^2 \bigl[\gamma r_+ - a \cos{(\theta)}\bigr] - a^3 \sin^2{(\theta)} \cos{(\theta)} \, \bigl[r_+ + \gamma a \cos{(\theta)}\bigr]\Bigr)\, \textnormal{d}\theta \\ \\
& \hspace{0.4cm} + \rho_{0 +}^2 \sin{(\theta)} \, \Biggl(\left[r_+ - a^2 \sin^2{(\theta)} \left(\alpha_+ + \frac{r_+}{\rho_+^2} \right)\right] \, \bigl[r_+ + \gamma a \cos{(\theta)}\bigr] - \frac{a^3 \sin^2{(\theta)} \cos{(\theta)}}{\rho_+^2} \, \bigl[\gamma r_+ - a \cos{(\theta)}\bigr]\Biggr) \, \textnormal{d}\tilde{\varphi}_+\Biggr] \\ \\
\sdpb{\hspace{0.04cm} A^{(3)}_{\hspace{0.03cm} \gamma}} & = \frac{1}{\sqrt{2} \, \rho_+^4} \, \Biggl[\Bigl(- r_+ \rho_{0 +}^2 \bigl[r_+ + \gamma a \cos{(\theta)}\bigr] + a^3 \sin^2{(\theta)} \cos{(\theta)} \, \bigl[- \gamma r_+ + a \cos{(\theta)}\bigr]\Bigr)\, \textnormal{d}\theta \\ \\
& \hspace{0.4cm} + \rho_{0 +}^2 \sin{(\theta)} \, \Biggl(\biggl[r_+ - a^2 \sin^2{(\theta)} \, \biggl(\alpha_+ + \frac{r_+}{\rho_+^2} \biggr)\biggr] \, \bigl[\gamma r_+ - a \cos{(\theta)}\bigr] + \frac{a^3 \sin^2{(\theta)} \cos{(\theta)}}{\rho_+^2} \, \bigl[r_+ + \gamma a \cos{(\theta)}\bigr]\Biggr) \, \textnormal{d}\tilde{\varphi}_+\Biggr] \, .
\end{split}
\end{equation}
The associated Ashtekar--Barbero curvature 
\begin{equation*}
F_{\gamma}^{(i)} := F^{(i)}(\boldsymbol{A}_{\gamma}) = \textnormal{d}A^{(i)}_{\gamma} + \frac{1}{2} \, \epsilon\indices{^{(i)}_{(j) (k)}} \, A^{(j)}_{\gamma} \wedge A^{(k)}_{\gamma} 
\end{equation*}
reads
\begin{equation}\label{ABCU}
\begin{split}
\sdpb{\hspace{0.04cm} F^{(1)}_{\hspace{0.04cm} \gamma}} & = 2 \, \Biggl[\textnormal{Re}{\bigl(\Psi_{2 \, | \, r = r_+}\bigr) + \gamma \, \textnormal{Im}{\bigl(\Psi_{2 \, | \, r = r_+}\bigr)} + \frac{1 + \gamma^2}{4 \rho_+^6} \, \biggl(r_+^4 - a^4 \cos^4{(\theta)} + \frac{a^2 \rho_{0 +}^2}{2} \, \bigl(3 \cos^2{(\theta)} - 1\bigr)\biggr)\Biggr]\, \sdpb{\hspace{0.04cm} \Sigma^{(1)}}} \\ \\
\sdpb{\hspace{0.04cm} F^{(2)}_{\hspace{0.04cm} \gamma}} & = \frac{3 a \sin{(\theta)}}{\sqrt{2} \, \rho_+^2} \, \Biggl[\bigl(- \gamma r_+ + a \cos{(\theta)}\bigr) \, \textnormal{Re}{\bigl(\Psi_{2 \, | \, r = r_+}\bigr)} + \bigl(r_+ + \gamma a \cos{(\theta)}\bigr) \, \textnormal{Im}{\bigl(\Psi_{2 \, | \, r = r_+}\bigr)} \\
& \hspace{2.1cm} + \frac{a \cos{(\theta)} \bigl(1 + \gamma^2\bigr)}{6 \rho_+^4} \, \bigl(5 r_+^2 - a^2 \cos{(2 \theta)}\bigr)\Biggr]\, \sdpb{\hspace{0.04cm} \Sigma^{(1)}} \\ \\
\sdpb{\hspace{0.04cm} F^{(3)}_{\hspace{0.04cm} \gamma}} & = \frac{3 a \sin{(\theta)}}{\sqrt{2} \, \rho_+^2} \, \Biggl[\bigl(r_+ + \gamma a \cos{(\theta)}\bigr) \, \textnormal{Re}{\bigl(\Psi_{2 \, | \, r = r_+}\bigr)} + \bigl(\gamma r_+ - a \cos{(\theta)}\bigr) \, \textnormal{Im}{\bigl(\Psi_{2 \, | \, r = r_+}\bigr)} \\
& \hspace{2.1cm} + \frac{r_+ \bigl(1 + \gamma^2\bigr)}{6 \rho_+^4} \, \Bigl(3 r_+^2 - a^2 \, \bigl[1 + 2 \cos{(2 \theta)}\bigr]\Bigr)\Biggr]\, \sdpb{\hspace{0.04cm} \Sigma^{(1)}} \, .
\end{split}
\end{equation}

\section{SU($2$) Chern--Simons Form of the Horizon Boundary Condition}

\noindent The pullback of the Ashtekar--Barbero curvature to the horizon $2$-spheres (\ref{ABCU}) is of the form 
\begin{equation} \label{ORIGFORM}
\sdpb{\hspace{0.04cm} F^{(i)}_{\hspace{0.04cm} \gamma}} = f^{(i)}(\theta) \, \sdpb{\hspace{0.04cm} \Sigma^{(1)}} \, ,
\end{equation} 
where $f^{(i)}(\theta): [0, \pi] \mapsto \mathbb{C}$, $i \in \{1, 2, 3\}$, are disparate complex-valued functions. This relation differs from the one of the Schwarzschild isolated horizon case for which the Ashtekar--Barbero curvature and the Plebanski $2$-form, both projected onto the horizon $2$-spheres, are proportional
\begin{equation}\label{SU2hbc}
\sdpb{\hspace{0.04cm} F^{(i)}_{\hspace{0.04cm} \gamma}} = \alpha_0  \sdpb{\hspace{0.04cm} \Sigma^{(i)}} \, , \,\,\,\,\,\, \alpha_0 = \textnormal{const.} 
\end{equation}
As the latter equation is the key constraint in the classical symplectic theory of Schwarzschild isolated horizons, relating the bulk geometry degrees of freedom to the isolated horizon surface degrees of freedom, and, furthermore, in their SU(2) Chern--Simons treatment in the framework of loop quantum gravity \cite{ENPP}, it is desirable to have such an equation satisfied also for Kerr isolated horizons. (The symplectic structure and a first quantum model of Kerr isolated horizons based on an SU($2$) Chern--Simons theory can be found in \cite{FPPR}.) In the following, we present a method for the construction of a new connection whose curvature satisfies this particular form of the horizon boundary condition.

Let $\sdpb{\hspace{0.04cm} \boldsymbol{F}_{\gamma}}$ be a curvature of the form (\ref{ORIGFORM}). We consider this curvature as a vector in $\mathbb{R}^3$ with three non-zero components, i.e., 
\begin{equation*}
\sdpb{\hspace{0.03cm} \boldsymbol{F}_{\gamma}} = \boldsymbol{f} \, \sdpb{\hspace{0.04cm} \Sigma^{(1)}} \, ,
\end{equation*}
where $\boldsymbol{f} = \boldsymbol{f}(\theta) = f^{(i)}(\theta) \, \widehat{\boldsymbol{e}}_{(i)} \in \mathbb{R}^3$ with $\widehat{\boldsymbol{e}}_{(i)}$ denoting the basis vectors along the $1, 2, 3$-directions, respectively. Note that the components of $\boldsymbol{f}$ can be directly read off from (\ref{ABCU}) and that the $2$-form $\sdpb{\hspace{0.04cm} \Sigma^{(1)}}$ acts merely as a multiplicative factor. Besides, this identification implies that $\gamma$ is restricted to values in $\mathbb{R} \backslash \{0\}$. Since the pullback of the Plebanski $2$-form to the horizon $2$-spheres $\sdpb{\hspace{0.04cm} \boldsymbol{\Sigma}}$, which we also treat as a vector in $\mathbb{R}^3$, has only one non-vanishing component, namely $\sdpb{\hspace{0.04cm} \Sigma^{(1)}} \not= 0$, we construct a new curvature $\sdpb{\hspace{0.04cm} \boldsymbol{\mathcal{F}}}_{\gamma}$ that fulfills the relation
\begin{equation} \label{IntermedCurv}
\sdpb{\hspace{0.04cm} \boldsymbol{\mathcal{F}}}_{\gamma} := R \, \sdpb{\hspace{0.04cm} \boldsymbol{F}_{\gamma}} = \|\boldsymbol{f}\| \, \widehat{\boldsymbol{e}}_{(1)} \, \sdpb{\hspace{0.04cm} \Sigma^{(1)}} = \|\boldsymbol{f}\| \, \sdpb{\hspace{0.04cm} \boldsymbol{\Sigma}} \, ,
\end{equation}
in which we align $\sdpb{\hspace{0.03cm} \boldsymbol{F}_{\gamma}}$ with the $1$-direction using an SO($3$) transformation $R$. We point out that the choice of rotation is not unique. Here, we use a rotation that transports the head of the vector $\sdpb{\hspace{0.04cm} \boldsymbol{F}_{\gamma}}$ along the shortest path between the initial and the end point on the enclosing sphere (with radius $\|\sdpb{\hspace{0.04cm} \boldsymbol{F}_{\gamma}}\|$). This is realized by rotating the curvature by an angle 
\begin{equation}\label{angle}
[0, \pi] \ni \alpha = \textnormal{arctan}\left(\frac{\sqrt{\bigl(f^{(2)}\bigr)^2 + \bigl(f^{(3)}\bigr)^2}}{f^{(1)}}\right) + \begin{cases} 0 & \, \textnormal{for} \,\,\,\,\, f^{(1)} \geq 0 \\  
\pi & \, \textnormal{for} \,\,\,\,\, f^{(1)} < 0 \end{cases} 
\end{equation}
around a (unit) normal to the $\sdpb{\hspace{0.04cm} \boldsymbol{F}_{\gamma}} - 1$ plane given by 
\begin{equation}\label{ndir}
\boldsymbol{n} = \frac{\boldsymbol{f} \times \|\boldsymbol{f}\| \, \widehat{\boldsymbol{e}}_{(1)}}{\big\|\boldsymbol{f} \times  \|\boldsymbol{f}\| \, \widehat{\boldsymbol{e}}_{(1)}\big\|} = \frac{1}{\sqrt{\bigl(f^{(2)}\bigr)^2 + \bigl(f^{(3)}\bigr)^2}} \, \left(\begin{array}{c}
0 \\
f^{(3)} \\
- f^{(2)}
\end{array}\right) \, .
\end{equation}
Applying this rotation to $\sdpb{\hspace{0.04cm} \boldsymbol{F}_{\gamma}}$, we get
\begin{equation} \label{vecftr}
\sdpb{\hspace{0.04cm} \boldsymbol{\mathcal{F}}}_{\gamma} = R_{\boldsymbol{n}}(\alpha) \, \boldsymbol{f} \, \sdpb{\hspace{0.04cm} \Sigma^{(1)}} = \Bigl[\bigl(\boldsymbol{n} \cdot \boldsymbol{f}\bigr) \boldsymbol{n} + \cos{(\alpha)} \, \bigl(\boldsymbol{n} \times \boldsymbol{f}\bigr) \times \boldsymbol{n} + \sin{(\alpha)} \, \bigl(\boldsymbol{n} \times \boldsymbol{f}\bigr)\Bigr] \, \sdpb{\hspace{0.04cm} \Sigma^{(1)}} = \Bigl[\cos{(\alpha)} \boldsymbol{f} + \sin{(\alpha)} \,  \bigl(\boldsymbol{n} \times \boldsymbol{f}\bigr)\Bigr] \, \sdpb{\hspace{0.04cm} \Sigma^{(1)}} \, .
\end{equation}
We now define the connection that corresponds to the new curvature (\ref{IntermedCurv}). For this purpose, we rotate the connection (\ref{ABCO}) employing the SU($2$) representation $\sdpb{\hspace{0.02cm} \boldsymbol{A}_{\hspace{0.03cm} \gamma}} = \textnormal{i}/2 \, \sdpb{\hspace{0.04cm} A^{(i)}_{\hspace{0.03cm} \gamma}} \sigma_{(i)}$ and the correspondence  
\begin{equation*}
R_{\boldsymbol{n}}(\alpha) = e^{\alpha \, \boldsymbol{n} \cdot \boldsymbol{J}} \,\,\,\,\, \longleftrightarrow \,\,\,\,\, G_{\boldsymbol{n}}(\alpha) = e^{\textnormal{i} \alpha \, \boldsymbol{n} \cdot \boldsymbol{\sigma}/2} = \textnormal{id}_{\mathbb{C}^2} \cos{\left(\frac{\alpha}{2}\right)} + \textnormal{i} \, (\boldsymbol{n} \cdot \boldsymbol{\sigma}) \sin{\left(\frac{\alpha}{2}\right)} \, ,
\end{equation*}
where the $J^i$ are the generators of SO($3$) and the $\sigma^i$ are the Pauli matrices. According to the transformation law 
\begin{equation*}
\sdpb{\hspace{0.02cm} \boldsymbol{\mathcal{A}}_{\hspace{0.03cm} \gamma}} = G_{\boldsymbol{n}}(\alpha) \, \sdpb{\hspace{0.02cm} \boldsymbol{A}_{\hspace{0.03cm} \gamma}} \, G_{\boldsymbol{n}}^{- 1}(\alpha) - \bigl(\textnormal{d} G_{\boldsymbol{n}}(\alpha)\bigr) \, G^{- 1}_{\boldsymbol{n}}(\alpha) \, ,
\end{equation*}
we find
\begin{equation*}
\sdpb{\hspace{0.04cm} \mathcal{A}^{(i)}_{\hspace{0.03cm} \gamma}} := \sdpb{\hspace{0.04cm} A^{(i)}_{\hspace{0.03cm} \gamma}} - \sin{(\alpha)} \, \biggl(\textnormal{d}n^i + \Bigl[\sdpb{\hspace{0.02cm} \boldsymbol{A}_{\hspace{0.03cm} \gamma}}, \boldsymbol{n}\Bigr]^i\biggr) + 2 \sin^2{\left(\frac{\alpha}{2}\right)} \, \biggl([\boldsymbol{n}, \textnormal{d}\boldsymbol{n}]^i - \Bigl(\sdpb{\hspace{0.02cm} \boldsymbol{A}_{\hspace{0.03cm} \gamma}} \cdot \boldsymbol{n}\Bigr) \, n^i\biggr) - n^i \, \textnormal{d}\alpha \, .
\end{equation*}
Substitution of the rotation angle (\ref{angle}) and the unit normal (\ref{ndir}) leads to the expressions
\begin{equation}\label{rotco}
\begin{split}
\sdpb{\hspace{0.04cm} \mathcal{A}^{(1)}_{\hspace{0.03cm} \gamma}} & = \sdpb{\hspace{0.04cm} A^{(1)}_{\hspace{0.03cm} \gamma}} + \frac{f^{(2)} \sdpb{\hspace{0.04cm} A^{(2)}_{\hspace{0.03cm} \gamma}} + f^{(3)} \sdpb{\hspace{0.04cm} A^{(3)}_{\hspace{0.03cm} \gamma}}}{\|\boldsymbol{f}\|} + \left(1 - \frac{f^{(1)}}{\|\boldsymbol{f}\|}\right) \, 
\frac{f^{(2)} \, \textnormal{d} f^{(3)} - f^{(3)} \, \textnormal{d} f^{(2)}}{\bigl(f^{(2)}\bigr)^2 + \bigl(f^{(3)}\bigr)^2} \\ \\
\sdpb{\hspace{0.04cm} \mathcal{A}^{(2)}_{\hspace{0.03cm} \gamma}} & = \sdpb{\hspace{0.04cm} A^{(2)}_{\hspace{0.03cm} \gamma}} - \frac{\textnormal{d}f^{(3)} + f^{(2)} \sdpb{\hspace{0.04cm} A^{(1)}_{\hspace{0.03cm} \gamma}}}{\|\boldsymbol{f}\|} + \frac{f^{(3)} \, \textnormal{d}f^{(1)}}{\|\boldsymbol{f}\|^2} \\ \\
& \hspace{1.07cm} + \left(1 -  \frac{f^{(1)}}{\|\boldsymbol{f}\|}\right) \, \frac{f^{(3)}}{\bigl(f^{(2)}\bigr)^2 + \bigl(f^{(3)}\bigr)^2} \, \biggl[\frac{f^{(2)} \, \textnormal{d}f^{(2)} + f^{(3)} \, \textnormal{d}f^{(3)}}{\|\boldsymbol{f}\|} + f^{(2)} \sdpb{\hspace{0.04cm} A^{(3)}_{\hspace{0.03cm} \gamma}} - f^{(3)} \sdpb{\hspace{0.04cm} A^{(2)}_{\hspace{0.03cm} \gamma}}\biggr] \\ \\
\sdpb{\hspace{0.04cm} \mathcal{A}^{(3)}_{\hspace{0.03cm} \gamma}} & = \sdpb{\hspace{0.04cm} A^{(3)}_{\hspace{0.03cm} \gamma}} + \frac{\textnormal{d}f^{(2)} - f^{(3)} \sdpb{\hspace{0.04cm} A^{(1)}_{\hspace{0.03cm} \gamma}}}{\|\boldsymbol{f}\|} - \frac{f^{(2)} \, \textnormal{d}f^{(1)}}{\|\boldsymbol{f}\|^2} \\ \\
& \hspace{1.07cm} - \left(1 -  \frac{f^{(1)}}{\|\boldsymbol{f}\|}\right) \, \frac{f^{(2)}}{\bigl(f^{(2)}\bigr)^2 + \bigl(f^{(3)}\bigr)^2} \, \biggl[\frac{f^{(2)} \, \textnormal{d}f^{(2)} + f^{(3)} \, \textnormal{d}f^{(3)}}{\|\boldsymbol{f}\|} + f^{(2)} \sdpb{\hspace{0.04cm} A^{(3)}_{\hspace{0.03cm} \gamma}} - f^{(3)} \sdpb{\hspace{0.04cm} A^{(2)}_{\hspace{0.03cm} \gamma}}\biggr] \, .
\end{split}
\end{equation}
The next step is to construct a curvature that satisfies the relation 
\begin{equation} \label{thirdcurv}
\sdpb{\hspace{0.04cm} \boldsymbol{\mathscrbf{F}}}_{\gamma} := \frac{\alpha_0}{ \|\boldsymbol{f}\|} \, \sdpb{\hspace{0.04cm} \boldsymbol{\mathcal{F}}}_{\gamma} = \alpha_0 \, \sdpb{\hspace{0.04cm} \boldsymbol{\Sigma}} \, .
\end{equation}
Therefore, to achieve the deformation $\|\boldsymbol{f}\| \rightarrow \alpha_0$, we use a simple rescaling transformation on $\sdpb{\hspace{0.04cm} \boldsymbol{\mathcal{F}}}_{\gamma}$. The associated connection $\sdpb{\hspace{0.00cm} \boldsymbol{\mathscrbf{A}}}_{\hspace{-0.04cm} \gamma}$
is obtained by the replacement 
\begin{equation*}
\tilde{\varphi}_+ \rightarrow \frac{\alpha_0}{\|\boldsymbol{f}\|} \, \tilde{\varphi}_+ 
\end{equation*}
that amounts to the substitution 
\begin{equation*}
\textnormal{d}\tilde{\varphi}_+ \rightarrow \frac{\alpha_0}{\|\boldsymbol{f}\|} \, \biggl(\textnormal{d}\tilde{\varphi}_+ - \frac{\tilde{\varphi}_+}{\|\boldsymbol{f}\|^2} \, \Bigl[f^{(1)} \, \textnormal{d}f^{(1)} + f^{(2)} \, \textnormal{d}f^{(2)} + f^{(3)} \, \textnormal{d}f^{(3)}\Bigr]\biggr) 
\end{equation*}
in the formulas given in (\ref{rotco}). A few concluding remarks are in order. The aim of this section was to find a connection, other than the Ashtekar--Barbero connection, with a curvature that is proportional to the Plebanski $2$-form on the Kerr isolated horizon $2$-spheres. To this end, we applied an SU(2) gauge transformation and a diffeomorphism solely to the Ashtekar--Barbero connection (not affecting the Plebanski $2$-form), in this way constructing a new horizon boundary connection with the desired property. As Eq.(\ref{SU2hbc}) is both gauge and diffeomorphism invariant, it is obvious that we did not use these transformations in order to turn Eq.(\ref{ORIGFORM}) into Eq.(\ref{SU2hbc}). The proposed method guarantees that the constructed quantity is still a connection. This can be directly seen from the fact that it is derived from an $\mathfrak{su}_{2}(\mathbb{C})$-valued $1$-form, namely $\sdpb{\hspace{0.02cm} \boldsymbol{A}_{\hspace{0.03cm} \gamma}}$, that transforms under SU(2) gauge transformations as a connection and under diffeomorphisms as a $1$-form, only using such kind of transformations according to the proper transformation laws. We remark in passing that the proposed strategy to obtain the desired horizon boundary condition is general in the sense that it works virtually on any $2$-surface in spacetime and, thus, it is not specific to the isolated horizon $2$-spheres under consideration.

\section{Slowly Rotating Kerr Isolated Horizons}

\noindent We present the first-order formalism of the Ashtekar--Barbero connection variables on the horizon $2$-spheres for small angular momenta $a/r_+ \ll 1$. Moreover, we explicitly compute the new connection $\sdpb{\hspace{0.00cm} \boldsymbol{\mathscrbf{A}}}_{\hspace{-0.04cm} \gamma}$ and the associated curvature $\sdpb{\hspace{0.04cm} \boldsymbol{\mathscrbf{F}}}_{\hspace{-0.04cm}\gamma}$, which satisfies the SU($2$) Chern--Simons form of the horizon boundary condition, according to the construction method derived in the previous section. Expanding the Ashtekar--Barbero connection (\ref{ABCO}) and the Ashtekar--Barbero curvature (\ref{ABCU}) up to the first order in $a/r_+$ yields 
\begin{equation}
\begin{split} \label{foabc}
\sdpb{\hspace{0.04cm} A^{(1)}_{\hspace{0.03cm} \gamma}} & \simeq - \frac{a \sin{(\theta)}}{r_+} \, \textnormal{d}\theta + \biggl(\cos{(\theta)} - \frac{3 \gamma a \sin^2{(\theta)}}{2 r_+}\biggr) \, \textnormal{d}\tilde{\varphi}_+ \\ \\
\sdpb{\hspace{0.04cm} A^{(2)}_{\hspace{0.03cm} \gamma}} & \simeq \frac{1}{\sqrt{2} \, r_+} \, \Bigl(\bigl[\gamma r_+ - a \cos{(\theta)}\bigr] \, \textnormal{d}\theta + \sin{(\theta)} \, \bigl[r_+ + \gamma a \cos{(\theta)}\bigr] \, \textnormal{d}\tilde{\varphi}_+\Bigr) \\ \\
\sdpb{\hspace{0.04cm} A^{(3)}_{\hspace{0.03cm} \gamma}} & \simeq \frac{1}{\sqrt{2} \, r_+} \, \Bigl(- \bigl[r_+ + \gamma a \cos{(\theta)}\bigr] \, \textnormal{d}\theta + \sin{(\theta)} \, \bigl[\gamma r_+ - a \cos{(\theta)}\bigr] \, \textnormal{d}\tilde{\varphi}_+\Bigr)
\end{split}
\end{equation}
and
\begin{equation} \label{foabcv}
\begin{split}
\sdpb{\hspace{0.04cm} F^{(1)}_{\hspace{0.04cm} \gamma}} & \simeq \biggl(\frac{\gamma^2 - 1}{2} - \frac{3 \gamma a \cos{(\theta)}}{r_+}\biggr) \sin{(\theta)} \,  \textnormal{d}\theta \wedge \textnormal{d}\tilde{\varphi}_+ = \frac{1}{r_+^2} \, \biggl(\frac{\gamma^2 - 1}{2} - \frac{3 \gamma a \cos{(\theta)}}{r_+}\biggr) \, \sdpb{\hspace{0.04cm} \Sigma^{(1)}_{\,\, 0}} \\ \\
\sdpb{\hspace{0.04cm} F^{(2)}_{\hspace{0.04cm} \gamma}} & \simeq \frac{1}{\gamma} \, \sdpb{\hspace{0.04cm} F^{(3)}_{\hspace{0.04cm} \gamma}} \simeq \frac{3 \gamma a \sin^2{(\theta)}}{2 \sqrt{2} \, r_+} \, \textnormal{d}\theta \wedge \textnormal{d}\tilde{\varphi}_+ = \frac{3 \gamma a \sin{(\theta)}}{2 \sqrt{2} \, r_+^3} \, \sdpb{\hspace{0.04cm} \Sigma^{(1)}_{\,\, 0}} \, ,
\end{split}
\end{equation}
where $\sdpb{\hspace{0.04cm} \Sigma^{(1)}_{\,\, 0}} := \sdpb{\hspace{0.04cm} \Sigma^{(1)}}(a/r_+ \ll 1) \simeq r_+^2 \sin{(\theta)} \, \textnormal{d}\theta \wedge \textnormal{d}\tilde{\varphi}_+$ is the first component of the first-order Plebanski $2$-form. Using (\ref{foabc}) and the components of the vector $\boldsymbol{f}$
\begin{equation*}
\begin{split}
& f^{(1)} \simeq \frac{1}{r_+^2} \, \biggl(\frac{\gamma^2 - 1}{2} - \frac{3 \gamma a \cos{(\theta)}}{r_+}\biggr) \\ \\
& f^{(2)} \simeq \frac{f^{(3)}}{\gamma} \simeq \frac{3 \gamma a \sin{(\theta)}}{2 \sqrt{2} \, r_+^3} \, ,
\end{split}
\end{equation*}
which are directly read off from (\ref{foabcv}), we find for the first-order connection (\ref{rotco})
\begin{equation} \label{foroco}
\begin{split}
\sdpb{\hspace{0.04cm} \mathcal{A}^{(1)}_{\hspace{0.03cm} \gamma}} & \simeq - \frac{a \sin{(\theta)}}{r_+} \, \textnormal{d}\theta + \biggl(\cos{(\theta)} - \frac{3 \gamma a \sin^2{(\theta)}}{r_+ (1 - \gamma^2)}\biggr) \, \textnormal{d}\tilde{\varphi}_+ \\ \\
\sdpb{\hspace{0.04cm} \mathcal{A}^{(2)}_{\hspace{0.03cm} \gamma}} & \simeq  \frac{1}{\sqrt{2} \, r_+} \, \Biggl(\left[\gamma r_+ - a \cos{(\theta)} \left(1 + \frac{3 \gamma^2}{\gamma^2 - 1}\right)\right] \textnormal{d}\theta + \sin{(\theta)} \,  \biggl[r_+ + \gamma a \cos{(\theta)} \, \biggl(1 - \frac{3}{\gamma^2 - 1}\biggr)\biggr] \, \textnormal{d}\tilde{\varphi}_+\Biggr) \\ \\
\sdpb{\hspace{0.04cm} \mathcal{A}^{(3)}_{\hspace{0.03cm} \gamma}} & \simeq  \frac{1}{\sqrt{2} \, r_+} \, \Biggl(- \biggl[r_+ + \gamma a \cos{(\theta)} \, \biggl(1 - \frac{3}{\gamma^2 - 1}\biggr)\biggr] \, \textnormal{d}\theta + \sin{(\theta)} \left[\gamma r_+ - a \cos{(\theta)} \left(1 + \frac{3 \gamma^2}{\gamma^2 - 1}\right)\right] \textnormal{d}\tilde{\varphi}_+\Biggr) \, .
\end{split}
\end{equation}
The corresponding first-order curvature (\ref{vecftr}) becomes
\begin{equation}\label{forocu}
\begin{split}
\sdpb{\hspace{0.04cm} \mathcal{F}^{(1)}_{\hspace{0.04cm} \gamma}} & \simeq \frac{1}{r_+^2} \, \biggl(\frac{\gamma^2 - 1}{2} - \frac{3 \gamma a \cos{(\theta)}}{r_+}\biggr) \, \sdpb{\hspace{0.05cm} \Sigma^{(1)}_{\,\, 0}} \\ \\
\sdpb{\hspace{0.04cm} \mathcal{F}^{(2)}_{\hspace{0.04cm} \gamma}} & \simeq \frac{1}{\gamma} \, \sdpb{\hspace{0.04cm} \mathcal{F}^{(3)}_{\hspace{0.04cm} \gamma}} = \mathcal{O}(a^2) \simeq 0 \, .
\end{split}
\end{equation}
Note that these expressions are restricted to values of the Barbero--Immirzi parameter for which both $\gamma^2 - 1 \gg a / r_+$ and $\gamma^2 - 1 \geq 12 \, a \, |\gamma \cos{(\theta)}| / r_+$ hold. The connection and curvature for the remaining values can be obtained similarly. Next, choosing the constant $\alpha_0$ such that it reproduces the value of the Schwarzschild isolated horizon case in the limit $a \searrow 0$ 
\begin{equation*}
\alpha_0 := \frac{\gamma^2 - 1}{2 r_+^2} + a \, \epsilon_0 \, ,
\end{equation*}
where $a \, \epsilon_0$ is a small first-order correction, the replacement for the calculation of the connection $\sdpb{\hspace{0.00cm} \boldsymbol{\mathscrbf{A}}}_{\hspace{-0.04cm} \gamma}$ and the curvature (\ref{thirdcurv}) reads
\begin{equation*}
\tilde{\varphi}_+ \rightarrow \Biggl(1 + \frac{2 a}{\gamma^2 - 1} \, \biggl[r_+^2 \epsilon_0 + \frac{3 \gamma \cos{(\theta)}}{r_+}\biggr]\Biggr) \, \tilde{\varphi}_+ \, .
\end{equation*}
Consequently, substituting 
\begin{equation*}
\textnormal{d}\tilde{\varphi}_+ \rightarrow \Biggl(1 + \frac{2 a}{\gamma^2 - 1} \, \biggl[r_+^2 \epsilon_0 + \frac{3 \gamma \cos{(\theta)}}{r_+}\biggr]\Biggr) \, \textnormal{d}\tilde{\varphi}_+ - \frac{6 \gamma a \sin{(\theta)}}{r_+ (\gamma^2 - 1)} \, \tilde{\varphi}_+ \, \textnormal{d}\theta
\end{equation*}
into (\ref{foroco}) and (\ref{forocu}) leads to 
\begin{equation*}
\begin{split}
\sdpb{\hspace{0.04cm} \mathscr{A}^{(1)}_{\hspace{0.03cm} \gamma}} & \simeq - \frac{a \sin{(\theta)}}{r_+} \, \biggl(1 + \frac{6 \gamma \cos{(\theta)}}{\gamma^2 - 1} \, \tilde{\varphi}_+\biggr) \, \textnormal{d}\theta + \Biggl(\cos{(\theta)} + \frac{a}{\gamma^2 - 1} \, \biggl[2 r_+^2 \epsilon_0 \cos{(\theta)} + \frac{3 \gamma}{r_+} \, \bigl(1 + \cos^2{(\theta)}\bigr)\biggr]\Biggr) \, \textnormal{d}\tilde{\varphi}_+ \\ \\
\sdpb{\hspace{0.04cm} \mathscr{A}^{(2)}_{\hspace{0.03cm} \gamma}} & \simeq  \frac{1}{\sqrt{2} \, r_+} \, \Biggl(\left[\gamma r_+ - \frac{a}{\gamma^2 - 1} \, \Bigl(\cos{(\theta)} \, \bigl[4 \gamma^2 - 1\bigr] + 6 \gamma \sin^2{(\theta)} \,  \tilde{\varphi}_+\Bigr)\right] \textnormal{d}\theta \\ \\
& \hspace{5.2cm} + \sin{(\theta)} \, \biggl[r_+ + \frac{a}{\gamma^2 - 1} \, \Bigl(\gamma \cos{(\theta)} \, \bigl[\gamma^2 + 2\bigr] + 2 r_+^3 \epsilon_0\Bigr)\biggr] \,  \textnormal{d}\tilde{\varphi}_+\Biggr) \\ \\
\sdpb{\hspace{0.04cm} \mathscr{A}^{(3)}_{\hspace{0.03cm} \gamma}} & \simeq  \frac{1}{\sqrt{2} \, r_+} \, \Biggl(- \left[r_+ + \frac{\gamma a}{\gamma^2 - 1} \, \Bigl(\cos{(\theta)} \, \bigl[\gamma^2 - 4\bigr] + 6 \gamma \sin^2{(\theta)} \, \tilde{\varphi}_+\Bigr)\right] \textnormal{d}\theta \\ \\
& \hspace{5.2cm} + \sin{(\theta)} \, \biggl[\gamma r_+ + \frac{a}{\gamma^2 - 1} \,  \Bigl(\cos{(\theta)} \, \bigl[2 \gamma^2 + 1\bigr] + 2 \gamma r_+^3 \epsilon_0\Bigr)\biggr] \, \textnormal{d}\tilde{\varphi}_+\Biggr)
\end{split}
\end{equation*}
and
\begin{equation*}
\begin{split}
\sdpb{\hspace{0.04cm} \mathscr{F}^{(1)}_{\hspace{0.04cm} \gamma}} & \simeq \biggl(\frac{\gamma^2 - 1}{2 r_+^2} + a \, \epsilon_0\biggr) \, \sdpb{\hspace{0.04cm} \Sigma^{(1)}_{\,\, 0}} \\ \\
\sdpb{\hspace{0.04cm} \mathscr{F}^{(2)}_{\hspace{0.04cm} \gamma}} & \simeq \frac{1}{\gamma} \, \sdpb{\hspace{0.04cm} \mathscr{F}^{(3)}_{\hspace{0.04cm} \gamma}} \simeq 0 \, .
\end{split}
\end{equation*}
Thus, we have shown that this particular curvature satisfies the first-order horizon boundary condition of the desired SU($2$) Chern--Simons form (\ref{SU2hbc})
\begin{equation*}
\sdpb{\hspace{0.04cm} \mathscr{F}^{(i)}_{\hspace{0.04cm} \gamma}} = \biggl(\frac{\gamma^2 - 1}{2 r_+^2} + a \, \epsilon_0\biggr) \, \sdpb{\hspace{0.04cm} \Sigma^{(i)}_{\,\, 0}} 
\end{equation*}
with $\sdpb{\hspace{0.04cm} \Sigma^{(1)}_{\,\, 0}}$ given below (\ref{foabcv}) and $\sdpb{\hspace{0.04cm} \Sigma^{(2)}_{\,\, 0}} = \sdpb{\hspace{0.04cm} \Sigma^{(3)}_{\,\, 0}} = 0$, which concludes the analysis of the first-order formalism.

\vspace{0.5cm} 
 
\section*{Acknowledgments}

\noindent I am grateful to Alejandro Perez, Felix Finster, Simone Murro, Guillaume Idelon--Riton, Ernesto Frodden, and Daniele Pranzetti for useful discussions and comments, and to Katharina Proksch, Horst Fichtner, and Florian Schuppan for a careful reading of the first version of this paper. Furthermore, I thank the anonymous referees for helpful and constructive comments. This research was supported by a DFG Research Fellowship.

\vspace{0.5cm}

\begin{appendix}

\section{The Newman--Penrose Formalism} \label{AA}

\noindent Let $F\mathfrak{M}$ and $F^{\star}\mathfrak{M}$ be a null frame bundle and its dual on $\mathfrak{M}$. On their sections, we introduce local tetrad and co-tetrad basis $(\boldsymbol{e}_{(a)})$ and $(\boldsymbol{e}^{(a)})$ with $a \in \{0, 1, 2, 3\}$. These are related to the basis vectors of the sections of the tangent and cotangent bundles $T\mathfrak{M}$ and $T^{\star}\mathfrak{M}$ via 
$\boldsymbol{e}_{(a)} = e\indices{^{\mu}_{(a)}} \, \boldsymbol{e}_{\mu}$ and $\boldsymbol{e}^{(a)} = e\indices{_{\mu}^{(a)}} \, \boldsymbol{e}^{\mu}$, where $e\indices{^{\mu}_{(a)}}$ is an invertible, linear $4 \times 4$ matrix-valued map from $T\mathfrak{M}$ to $F\mathfrak{M}$. The tetrad and its dual satisfy the orthonormality condition $\boldsymbol{e}^{(a)}(\boldsymbol{e}_{(b)}) = \delta^{(a)}_{(b)}$ as well as the metric condition $\boldsymbol{g}(\boldsymbol{e}_{(a)}, \boldsymbol{e}_{(b)}) = \eta_{(a) (b)}$ with the constant, symmetric matrix $\eta_{(a) (b)}$. In the Newman--Penrose formalism \cite{NP, NP2}, the null tetrad consists of two real-valued vectors $\boldsymbol{l} = \boldsymbol{e}_{(0)} = \boldsymbol{e}^{(1)}$ and $\boldsymbol{n} = \boldsymbol{e}_{(1)} = \boldsymbol{e}^{(0)}$, as well as the complex-conjugated vectors $\boldsymbol{m} = \boldsymbol{e}_{(2)} = - \boldsymbol{e}^{(3)}$ and $\overline{\boldsymbol{m}} = \boldsymbol{e}_{(3)} = - \boldsymbol{e}^{(2)}$. These fulfill the null conditions 
\begin{equation*}
\boldsymbol{l} \cdot \boldsymbol{l} = \boldsymbol{n} \cdot \boldsymbol{n} = \boldsymbol{m} \cdot \boldsymbol{m} = \overline{\boldsymbol{m}} \cdot \overline{\boldsymbol{m}} = 0 \, , 
\end{equation*} 
the orthogonality conditions
\begin{equation*}
\boldsymbol{l} \cdot \boldsymbol{m} = \boldsymbol{l} \cdot \overline{\boldsymbol{m}} = \boldsymbol{n} \cdot \boldsymbol{m} = \boldsymbol{n} \cdot \overline{\boldsymbol{m}} = 0 \, ,
\end{equation*} 
and the cross-normalization conditions
\begin{equation*}
\boldsymbol{l} \cdot \boldsymbol{n} = - \, \boldsymbol{m} \cdot \overline{\boldsymbol{m}} = 1 \, ,
\end{equation*} 
which depend on the chosen signature convention. In this particular null frame, the metric is given by
\begin{equation*}
\boldsymbol{g} = \eta_{(a) (b)} \, \boldsymbol{e}^{(a)} \otimes \boldsymbol{e}^{(b)} = \boldsymbol{l} \otimes \boldsymbol{n} + \boldsymbol{n} \otimes \boldsymbol{l} - \boldsymbol{m} \otimes \overline{\boldsymbol{m}} - \overline{\boldsymbol{m}} \otimes \boldsymbol{m} \, ,
\end{equation*}
where 
\begin{equation} \label{locNPmetric}
\bigl(\eta_{(a) (b)}\bigr) = \begin{pmatrix} 0 & 1 & 0 & 0 \\ 1 & 0 & 0 & 0 \\ 0 & 0 & 0 & -1 \\ 0 & 0 & - 1 & 0 \end{pmatrix} \, .
\end{equation}
To determine the spin connection $\boldsymbol{\omega}$, one solves the first Maurer--Cartan equation of structure 
\begin{equation*}
\textnormal{d}\boldsymbol{e}\indices{^{\mu}} + \omega\indices{^{\mu}_{\nu}} \wedge \boldsymbol{e}\indices{^{\nu}} = 0 \, ,
\end{equation*}
which in the tetrad formulation reads
\begin{equation*}
\textnormal{d}\boldsymbol{e}\indices{^{(a)}} = \gamma\indices{^{(a)}_{(b) (c)}} \, \boldsymbol{e}\indices{^{(b)}} \wedge \boldsymbol{e}\indices{^{(c)}}
\end{equation*}
with the Ricci rotation coefficients 
\begin{equation*}
\gamma\indices{^{(a)}_{(b) (c)}} \, \boldsymbol{e}\indices{^{(b)}} = e\indices{_{\mu}^{(a)}} \, \textnormal{d}e\indices{^{\mu}_{(c)}} + e\indices{_{\mu}^{(a)}} \, e\indices{^{\nu}_{(c)}} \, \omega\indices{^{\mu}_{\nu}} \, .
\end{equation*}
In the Newman--Penrose formalism, the Ricci rotation coefficients are represented by  twelve special symbols called spin coefficients 
\begin{eqnarray} \label{rrcspc}
\kappa = \gamma\indices{_{(2) (0) (0)}}  & \,\,\,\, \varrho = \gamma\indices{_{(2) (0) (3)}} & \,\,\,\, \epsilon = \tfrac{1}{2} \bigl(\gamma\indices{_{(1) (0) (0)}} + \gamma\indices{_{(2) (3) (0)}}\bigr) \nonumber \\
\sigma = \gamma\indices{_{(2) (0) (2)}} & \,\,\,\, \mu = \gamma\indices{_{(1) (3) (2)}} & \,\,\,\, \gamma = \tfrac{1}{2} \bigl(\gamma\indices{_{(1) (0) (1)}} + \gamma\indices{_{(2) (3) (1)}}\bigr) \nonumber \\
\lambda = \gamma\indices{_{(1) (3) (3)}} & \,\,\,\, \tau = \gamma\indices{_{(2) (0) (1)}} & \,\,\,\, \alpha = \tfrac{1}{2} \bigl(\gamma\indices{_{(1) (0) (3)}} + \gamma\indices{_{(2) (3) (3)}}\bigr) \\
\nu = \gamma\indices{_{(1) (3) (1)}} & \,\,\,\, \pi = \gamma\indices{_{(1) (3) (0)}} & \,\,\,\, \beta = \tfrac{1}{2} \bigl(\gamma\indices{_{(1) (0) (2)}} + \gamma\indices{_{(2) (3) (2)}}\bigr) \, .\nonumber 
\end{eqnarray}
Accordingly, we obtain the first Maurer--Cartan equation of structure in the following form 
\begin{equation} \label{NPMCE}
\begin{split}
\textnormal{d}\boldsymbol{l} & = 2 \, \textnormal{Re}{(\epsilon)} \, \boldsymbol{n} \wedge \boldsymbol{l} - 2 \, \boldsymbol{n} \wedge \textnormal{Re}{(\kappa \, \overline{\boldsymbol{m}})} - 2 \,  \boldsymbol{l} \wedge \textnormal{Re}{\bigl([\tau - \overline{\alpha} - \beta] \, \overline{\boldsymbol{m}}\bigr)} + 2 \, \textnormal{i} \, \textnormal{Im}{(\varrho)} \, \boldsymbol{m} \wedge \overline{\boldsymbol{m}} \\ \\ 
\textnormal{d}\boldsymbol{n} & = 2 \, \textnormal{Re}{(\gamma)} \, \boldsymbol{n} \wedge \boldsymbol{l} - 2 \, \boldsymbol{n} \wedge \textnormal{Re}{\bigl([\overline{\alpha} + \beta - \overline{\pi}] \, \overline{\boldsymbol{m}}\bigr)} + 2 \, \boldsymbol{l} \wedge \textnormal{Re}{(\overline{\nu} \, \overline{\boldsymbol{m}})} + 2 \, \textnormal{i} \, \textnormal{Im}{(\mu)} \, \boldsymbol{m} \wedge \overline{\boldsymbol{m}} \\ \\
\textnormal{d}\boldsymbol{m} & = \overline{\textnormal{d}\overline{\boldsymbol{m}}} = (\overline{\pi} + \tau) \, \boldsymbol{n} \wedge \boldsymbol{l} + \bigl(2 \, \textnormal{i} \, \textnormal{Im}{(\epsilon)} - \varrho\bigr) \, \boldsymbol{n} \wedge \boldsymbol{m} - \sigma \,  \boldsymbol{n} \wedge \overline{\boldsymbol{m}} + \bigl(\overline{\mu} + 2 \, \textnormal{i} \, \textnormal{Im}{(\gamma)}\bigr) \, \boldsymbol{l} \wedge \boldsymbol{m} + \overline{\lambda} \, \boldsymbol{l} \wedge \overline{\boldsymbol{m}} - (\overline{\alpha} - \beta) \, \boldsymbol{m} \wedge \overline{\boldsymbol{m}} \, . 
\end{split}
\end{equation}
Finally, we specify the relevant classes of local Lorentz transformations that are applied to the tetrad and the spin coefficients. With the parameters $\xi, \psi \in \mathbb{R}$ and $\textsf{a}, \textsf{b} \in \mathbb{C}$, which are functions of the spacetime coordinates $(x^{\mu})$, these yield \cite{ChandraBook}
\begin{align} \label{LLTT13}
& \underline{\textnormal{Class I}:} & & \underline{\textnormal{Class III}:} \nonumber \\  
& \boldsymbol{l} \mapsto \boldsymbol{l'} = \boldsymbol{l} & & \boldsymbol{l} \mapsto \boldsymbol{l'} = \xi \, \boldsymbol{l} \nonumber \\
& \boldsymbol{n} \mapsto \boldsymbol{n'} = \boldsymbol{n} + |\textsf{a}|^2 \boldsymbol{l} + \overline{\textsf{a}} \, \boldsymbol{m} +\textsf{a} \, \overline{\boldsymbol{m}} & & \boldsymbol{n} \mapsto \boldsymbol{n'} = \xi^{-1} \boldsymbol{n} \\ 
& \boldsymbol{m} \mapsto \boldsymbol{m'} = \boldsymbol{m} + \textsf{a} \, \boldsymbol{l} & & \boldsymbol{m} \mapsto \boldsymbol{m'} = e^{\textnormal{i} \psi} \, \boldsymbol{m} \nonumber \\
& \overline{\boldsymbol{m}} \mapsto \overline{\boldsymbol{m}}{\, '} = \overline{\boldsymbol{m}} + \overline{\textsf{a}} \, \boldsymbol{l} & & \overline{\boldsymbol{m}} \mapsto \overline{\boldsymbol{m}}{\, '} = e^{-  \textnormal{i} \psi} \, \overline{\boldsymbol{m}} \nonumber
\end{align}
and
\begin{align} \label{LLTSC1}
& \underline{\textnormal{Class I}:} \nonumber \\
& \kappa \mapsto \kappa' = \kappa \, , \,\,\,\,\,\, \sigma \mapsto \sigma' = \sigma + \textsf{a} \, \kappa \, , \,\,\,\,\,\, \varrho \mapsto \varrho' = \varrho + \overline{\textsf{a}} \, \kappa \, , \,\,\,\,\,\, \epsilon \mapsto \epsilon' = \epsilon + \overline{\textsf{a}} \, \kappa \, , \,\,\,\,\,\, \tau \mapsto \tau' = \tau + \textsf{a} \, \varrho + \overline{\textsf{a}} \, \sigma + |a|^2 \kappa \, , \nonumber \\ 
& \gamma \mapsto \gamma' = \gamma + \textsf{a} \, \alpha + \overline{\textsf{a}} \, (\beta + \tau) + |\textsf{a}|^2 (\varrho + \epsilon) + \overline{\textsf{a}}^2 \sigma + \overline{\textsf{a}} \, |\textsf{a}|^2 \kappa \, , \,\,\,\,\,\, \pi \mapsto \pi' = \pi + 2 \, \overline{\textsf{a}} \, \epsilon + \overline{\textsf{a}}^2 \kappa + \boldsymbol{l} \, \overline{\textsf{a}} \, , \nonumber \\ 
& \lambda \mapsto \lambda' = \lambda + \overline{\textsf{a}} \, (2 \, \alpha + \pi) + \overline{\textsf{a}}^2 (\varrho + 2 \, \epsilon) + \overline{\textsf{a}}^3 \kappa + \overline{\boldsymbol{m}} \, \overline{\textsf{a}} + \overline{\textsf{a}} \, \boldsymbol{l} \, \overline{\textsf{a}} \, , \,\,\,\,\,\, \alpha \mapsto \alpha' = \alpha + \overline{\textsf{a}} \, (\varrho + \epsilon) + \overline{\textsf{a}}^2 \kappa \, , \\ 
& \mu \mapsto \mu' = \mu + \textsf{a} \, \pi + 2 \, \overline{\textsf{a}} \, \beta + 2 \, |\textsf{a}|^2 \epsilon + \overline{\textsf{a}}^2 \sigma + \overline{\textsf{a}} \, |\textsf{a}|^2 \kappa + \boldsymbol{m} \, \overline{\textsf{a}} + \textsf{a} \, \boldsymbol{l} \, \overline{\textsf{a}} \, , \,\,\,\,\,\, \beta \mapsto \beta' = \beta + \textsf{a} \, \epsilon + \overline{\textsf{a}} \, \sigma + |\textsf{a}|^2 \kappa \, , \nonumber \\ 
& \nu \mapsto \nu' = \nu + \textsf{a} \, \lambda + \overline{\textsf{a}} \, (\mu + 2 \, \gamma) + \overline{\textsf{a}}^2 (\tau + 2 \, \beta) + \overline{\textsf{a}}^3 \sigma + |\textsf{a}|^2 (\pi + 2 \, \alpha) + \overline{\textsf{a}} \, |\textsf{a}|^2 (\varrho + 2 \, \epsilon) + \overline{\textsf{a}}^2 |\textsf{a}|^2 \kappa + (|\textsf{a}|^2 \boldsymbol{l} + \boldsymbol{n} + \overline{\textsf{a}} \, \boldsymbol{m} + \textsf{a} \, \overline{\boldsymbol{m}}) \, \overline{\textsf{a}} \nonumber
\end{align}
\begin{align} \label{LLTSC3}
& \underline{\textnormal{Class III}:} \nonumber \\ 
& \kappa \mapsto \kappa' = \xi^2 \, e^{\textnormal{i} \psi} \, \kappa \, , \,\,\,\,\,\, \sigma \mapsto \sigma' = \xi \, e^{2 \textnormal{i} \psi} \, \sigma \, , \,\,\,\,\,\, \tau \mapsto \tau' =e^{\textnormal{i} \psi} \, \tau \, , \,\,\,\,\,\, \pi \mapsto \pi' = e^{- \textnormal{i} \psi} \, \pi \, , \,\,\,\,\,\, \varrho \mapsto \varrho' = \xi \varrho \, , \,\,\,\,\,\, \mu \mapsto \mu' = \xi^{- 1} \mu \, ,\nonumber \\ 
& \alpha \mapsto \alpha' = e^{- \textnormal{i} \psi} \, \alpha + \tfrac{\textnormal{i}}{2} \, e^{- \textnormal{i} \psi} \, \overline{\boldsymbol{m}} \, \psi + \tfrac{1}{2} \, \xi^{- 1} \, e^{- \textnormal{i} \psi} \, \overline{\boldsymbol{m}} \, \xi \, , \,\,\,\,\,\, \beta \mapsto \beta' = e^{\textnormal{i} \psi} \, \beta + \tfrac{\textnormal{i}}{2} \, e^{\textnormal{i} \psi} \, \boldsymbol{m} \, \psi + \tfrac{1}{2} \, \xi^{- 1} \,  e^{\textnormal{i} \psi} \, \boldsymbol{m} \, \xi \, , \\
& \gamma \mapsto \gamma' = \xi^{- 1} \gamma + \tfrac{1}{2} \, \xi^{- 2} \boldsymbol{n} \, \xi + \tfrac{\textnormal{i}}{2} \, \xi^{- 1} \boldsymbol{n} \, \psi \, , \,\,\,\,\,\, \epsilon \mapsto \epsilon' = \xi \epsilon + \tfrac{1}{2} \, \boldsymbol{l} \, \xi + \tfrac{\textnormal{i}}{2} \, \xi \, \boldsymbol{l} \, \psi \, , \,\,\,\,\,\, \lambda \mapsto \lambda' = \xi^{- 1} \, e^{- 2 \textnormal{i} \psi} \, \lambda \, ,  \,\,\,\,\,\, \nu \mapsto \nu' = \xi^{- 2} \, e^{- \textnormal{i} \psi} \, \nu \, . \nonumber
\end{align}

\section{Analytical Extension of Kerr Geometry --- From Boyer--Lindquist to Kruskal--Szekeres-like Coordinates} \label{AA2}

\noindent The Kerr metric $\boldsymbol{g}$ in Boyer--Lindquist coordinates $(t, r, \theta, \varphi)$, with $t \in \mathbb{R}$, $r \in \mathbb{R}_{> 0}$, $\theta \in \left[0, \pi\right]$, and $\varphi \in [0, 2 \pi)$, as well as with the signature convention $(+, -, -, -)$, is given by \cite{Kerr, BL} 
\begin{equation}\label{KerrM}
\begin{split}
\boldsymbol{g} & = \frac{\Delta}{\rho^2} \, \bigl(\textnormal{d}t - a \sin^2{(\theta)} \, \textnormal{d}\varphi\bigr) \otimes \bigl(\textnormal{d}t - a \sin^2{(\theta)} \, \textnormal{d}\varphi\bigr) - \frac{\sin^2{(\theta)}}{\rho^2} \, \bigl(\rho_0^2 \,  \textnormal{d}\varphi - a \, \textnormal{d}t\bigr) \otimes \bigl(\rho_0^2 \, \textnormal{d}\varphi - a \, \textnormal{d}t\bigr) - \frac{\rho^2}{\Delta} \,  \textnormal{d}r \otimes \textnormal{d}r - \rho^2 \, \textnormal{d}\theta \otimes \textnormal{d}\theta \, ,
\end{split}
\end{equation}
where $\Delta = \Delta(r) := (r - r_+) (r - r_-) = r^2 - 2 M r + a^2$ is the horizon function, $\rho^2 = \rho^2(r, \theta) := r^2 + a^2 \cos^2{(\theta})$, and $\rho_0^2 = \rho_0^2(r) := r^2 + a^2$. The parameter $M$ denotes the mass, $a M$ the angular momentum, and $r_{\pm} := M \pm \sqrt{M^2 - a^2}$ the event horizon and the Cauchy horizon of the Kerr black hole. For the analysis of Kerr isolated horizons, we require a coordinate system that is regular at the event horizon. As Boyer--Lindquist coordinates are singular there, they are not suitable for the purpose at hand. Instead, we apply horizon-penetrating Kruskal--Szekeres-like coordinates $(K, L, \theta, \tilde{\varphi}_+)$, with $K \in \mathbb{R}$, $L \in \mathbb{R}_{> 0}$, $\theta \in \left[0, \pi\right]$, and $\widetilde{\varphi}_+ \in [0, 2 \pi)$, which read \cite{KRUS}
\begin{equation} \label{KZCS}
K := e^{\alpha_+ r_{\star}} \sinh{(\alpha_+ t)} \,, \,\,\,\,\, L := e^{\alpha_+ r_{\star}} \cosh{(\alpha_+ t)} \,, \,\,\,\,\, \theta = \theta \, , \,\,\,\,\, \textnormal{and} \,\,\,\,\,\, \tilde{\varphi}_+ := \varphi - \frac{a t}{\rho^2_{0 +}} \, ,
\end{equation}
where 
\begin{equation*}
r_{\star} := r + \frac{r_+^2 + a^2}{r_+ - r_-} \ln{|r - r_+|} - \frac{r_-^2 + a^2}{r_+ - r_-} \ln{|r - r_-|}
\end{equation*}
is the Regge--Wheeler coordinate, $\alpha_+ := (r_+ - r_-)/(2 \rho^2_{0 +})$, and $\rho^2_{0 +} := \rho^2_0(r = r_+)$. Note that these particular coordinates do not provide regularity at the Cauchy horizon. Also, we restrict our study to the exterior Kerr geometry including the event horizon $r_+ \leq r < \infty$. The Boyer--Lindquist basis vectors $(\partial_t, \partial_r, \partial_{\theta}, \partial_{\varphi})$ and basis $1$-forms $(\textnormal{d}t, \textnormal{d}r, \textnormal{d}\theta, \textnormal{d}\varphi)$ transform according to
\begin{equation} \label{BLBV}
\partial_t = \alpha_+ \bigl(L \, \partial_K + K \, \partial_L\bigr) - \frac{a}{\rho_{0 +}^2} \, \partial_{\tilde{\varphi}_+} \, , \,\,\,\,\,\, \partial_r =  \frac{\alpha_+ \beta \, \rho_0^2}{L^2 - K^2} \, \bigl(K \, \partial_K + L \, \partial_L\bigr) \, ,\,\,\,\,\,\, \partial_{\theta} = \partial_{\theta} \, , \,\,\,\,\,\, \partial_{\varphi} = \partial_{\tilde{\varphi}_+} 
\end{equation} 
and
\begin{equation} \label{BL1F}
\textnormal{d}t = \frac{L \, \textnormal{d}K - K \, \textnormal{d}L}{\alpha_+ (L^2 - K^2)} \, ,\,\,\,\,\,\, \textnormal{d}r = \frac{L \, \textnormal{d}L - K \, \textnormal{d}K}{\alpha_+ \beta \, \rho_{0}^2} \, ,\,\,\,\,\,\, \textnormal{d}\theta = \textnormal{d}\theta \, , \,\,\,\,\,\, \textnormal{d}\varphi = \textnormal{d}\tilde{\varphi}_+ + \frac{a \, (L \, \textnormal{d}K - K \, \textnormal{d}L)}{\alpha_+ \rho_{0 +}^2 (L^2 - K^2)} \, ,  
\end{equation} 
respectively. In terms of the Kruskal--Szekeres-like coordinates (\ref{KZCS}), the Kerr metric (\ref{KerrM}) becomes
\begin{equation} \label{KerrmetrKSC}
\begin{split}
\boldsymbol{g} & = \Omega^2 \, \biggl(\frac{\rho^4}{\rho_0^4} \, \bigl(\textnormal{d}K \otimes \textnormal{d}K - \textnormal{d}L \otimes \textnormal{d}L\bigr) - \chi_1 \,  \bigl(L \, \textnormal{d}K - K \, \textnormal{d}L\bigr) \otimes \bigl(L \, \textnormal{d}K - K \, \textnormal{d}L\bigr) - \chi_2 \, \bigl(L \, \textnormal{d}K - K \, \textnormal{d}L\bigr) \otimes \textnormal{d} \tilde{\varphi}_+\biggr) \\ \\
& \hspace{0.4cm} - \rho^2 \, \textnormal{d}\theta \otimes \textnormal{d}\theta - \frac{\sin^2{(\theta)}}{\rho^2} \, \left(\rho_0^4 - a^2 \sin^2{(\theta)} \, \Delta\right) \textnormal{d}\tilde{\varphi}_+ \otimes \textnormal{d}\tilde{\varphi}_+
\end{split}
\end{equation}
with the functions
\begin{equation*}
\begin{split}
\Omega & := \frac{1}{\alpha_+ \sqrt{\beta} \, \rho} \, , \,\,\,\,\, \chi_1 := \frac{a^2 \sin^2{(\theta)}}{\beta \, \rho_{0 +}^4} \, \frac{r + r_+}{r - r_-} \, \Biggl(\frac{r + r_+}{r - r_-} + \frac{2 \rho_+^2}{\rho_0^2} \, \biggl[1 + \frac{a^2 \sin^2{(\theta)} \, (r^2 - r_+^2)}{2 \rho_0^2 \, \rho_+^2} \biggr]\Biggr) \, , \\ \\
\chi_2 & := \frac{2 a \alpha_+ \sin^2{(\theta)}}{\rho_{0 +}^2} \left(\rho_0^2 \, \frac{r + r_+}{r - r_-} + \rho_+^2\right) \, , \,\,\,\,\, \textnormal{and} \,\,\,\,\, \beta := \frac{e^{2 \alpha_+ r}}{(r - r_-)^{1 + (r_-^2 + a^2)/\rho_{0 +}^2}} \, .
\end{split}
\end{equation*}  
The event horizon is determined via the implicit equation
\begin{equation*}
L^2 - K^2 = \Delta \, \beta \, ,
\end{equation*}
which relates the Kruskal--Szekeres-like coordinates $K$ and $L$ to the radial Boyer--Lindquist coordinate $r$. Evaluating this equation at $r = r_+$, we obtain the two solutions $K = \pm L$. But since we consider black hole geometries, only the positive solution $K = L$ describing future event horizons is of interest.

\section{Regular Kinnersley Tetrad and Spin Coefficients in Kruskal--Szekeres-like Coordinates} \label{AA3}

\noindent The Kinnersley tetrad in Boyer--Lindquist coordinates yields \cite{Kinn, ChandraBook}
\begin{equation}\label{BLtetradvectors}
\begin{split}
\boldsymbol{l} & = \frac{1}{\Delta} \left(\rho_0^2 \, \partial_t + \Delta \, \partial_r + a \, \partial_{\varphi}\right) \\ \\
\boldsymbol{n} & = \frac{1}{2 \rho^2} \left(\rho_0^2 \, \partial_{t} - \Delta \, \partial_r + a \, \partial_{\varphi}\right) \\ \\
\boldsymbol{m} & = \frac{1}{\sqrt{2} \, \overline{\varsigma}} \bigl(\textnormal{i} a \sin{(\theta)} \, \partial_t - \partial_{\theta} + \textnormal{i} \csc{(\theta)} \,  \partial_{\varphi}\bigr)\\ \\
\overline{\boldsymbol{m}} & = - \frac{1}{\sqrt{2} \, \varsigma} \, \bigl(\textnormal{i} a \sin{(\theta)} \, \partial_t + \partial_{\theta} + \textnormal{i} \csc{(\theta)} \, \partial_{\varphi}\bigr) 
\end{split}
\end{equation}
with the quantity $\varsigma = \varsigma(r, \theta) := r + \textnormal{i} a \cos{(\theta)}$ satisfying $\varsigma \, \overline{\varsigma} = \rho^2$. The corresponding dual Kinnersley co-tetrad is given by
\begin{equation}\label{BLtetrad}
\begin{split}
\boldsymbol{l} & = \textnormal{d}t - \frac{\rho^2}{\Delta} \, \textnormal{d}r - a \sin^2{(\theta)} \, \textnormal{d}\varphi \\ \\
\boldsymbol{n} & = \frac{\Delta}{2 \rho^2} \, \left(\textnormal{d}t + \frac{\rho^2}{\Delta} \, \textnormal{d}r - a \sin^2{(\theta)} \, \textnormal{d}\varphi\right) \\ \\
\boldsymbol{m} & = \frac{1}{\sqrt{2} \, \overline{\varsigma}} \, \left(\textnormal{i} a \sin{(\theta)} \, \textnormal{d}t + \rho^2 \, \textnormal{d}\theta - \textnormal{i} \rho_0^2 \sin{(\theta)} \, \textnormal{d}\varphi\right) \\ \\
\overline{\boldsymbol{m}} & = \frac{1}{\sqrt{2} \, \varsigma} \, \left(- \textnormal{i} a \sin{(\theta)} \, \textnormal{d}t + \rho^2 \, \textnormal{d}\theta + \textnormal{i} \rho_0^2 \sin{(\theta)} \, \textnormal{d}\varphi\right) \, .
\end{split}
\end{equation}
Substituting the basis vectors and $1$-forms (\ref{BLBV}) and (\ref{BL1F}) into (\ref{BLtetradvectors}) and (\ref{BLtetrad}), we obtain the Kinnersley tetrad and its dual in Kruskal--Szekeres-like coordinates. Although the underlying coordinate system is now regular up to the Cauchy horizon, the resulting expressions for the frame vector and $1$-form $\boldsymbol{l}$ are still singular at the event horizon. As these singularities arise in multiplicative factors, they can be eliminated by rescaling, using a class III local Lorentz transformation (\ref{LLTT13}) with parameters
\begin{equation*}
\xi = \frac{L - K}{\sqrt{2 \beta} \, \rho} \,\,\,\,\,\, \textnormal{and} \,\,\,\,\,\, \psi = 0 \, .
\end{equation*}
Applying this transformation leads to a Kinnersley tetrad in Kruskal--Szekeres-like coordinates that is regular at the event horizon 
\begin{equation}\label{NPBV}
\begin{split}
\boldsymbol{l}' & = \frac{\rho_0^2}{\sqrt{2} \, \rho^2 \Omega} \biggl(\partial_K + \partial_L - \frac{a \, (L - K)}{\alpha_+ \beta \, \rho_{0 +}^2 \rho_0^2} \, \frac{r + r_+}{r - r_-} \, \partial_{\tilde{\varphi}_+}\biggr)\\ \\
\boldsymbol{n}' & = \frac{\rho_0^2}{\sqrt{2} \, \rho^2 \Omega} \biggl(\partial_K - \partial_L - \frac{a \, (L + K)}{\alpha_+ \beta \, \rho_{0 +}^2 \rho_0^2} \, \frac{r + r_+}{r - r_-} \, \partial_{\tilde{\varphi}_+}\biggr)\\ \\
\boldsymbol{m}' & = \frac{1}{\sqrt{2} \, \overline{\varsigma}} \biggl(\textnormal{i} a \sin{(\theta)} \, \alpha_+ \bigl[L \, \partial_K + K \, \partial_L\bigr] - \partial_{\theta} + \frac{\textnormal{i} \rho_+^2}{\rho^2_{0 +} \sin{(\theta)}} \,  \partial_{\tilde{\varphi}_+}\biggr)\\ \\
\overline{\boldsymbol{m}}{\, '} & = - \frac{1}{\sqrt{2} \, \varsigma} \biggl(\textnormal{i} a \sin{(\theta)} \, \alpha_+ \bigl[L \, \partial_K + K \, \partial_L\bigr] + \partial_{\theta} + \frac{\textnormal{i} \rho_+^2}{\rho^2_{0 +} \sin{(\theta)}} \,  \partial_{\tilde{\varphi}_+}\biggr) 
\end{split}
\end{equation}
and to a regular dual co-tetrad 
\begin{equation}\label{NPZ}
\begin{split}
\boldsymbol{l}' & = \frac{\Omega}{\sqrt{2}} \, \Biggl(\left[\frac{\rho_+^2}{\rho_{0 +}^2} + \frac{a^2 \sin^2{(\theta)} \, K \, (L - K)}{\rho_{0 +}^2 \rho_0^2 \, \beta} \, \frac{r + r_+}{r - r_-}\right] \textnormal{d}K - \left[\frac{\rho_+^2}{\rho_{0 +}^2} + \frac{a^2 \sin^2{(\theta)} \, L \, (L - K)}{\rho_{0 +}^2 \rho_0^2 \, \beta} \, \frac{r + r_+}{r - r_-}\right] \textnormal{d}L \\
& \hspace{1.3cm} - a \sin^2{(\theta)} \, \alpha_+ (L - K) \, \textnormal{d}\tilde{\varphi}_+\Biggr) \\ \\
\boldsymbol{n}' & = \frac{\Omega}{\sqrt{2}} \, \Biggl(\left[\frac{\rho_+^2}{\rho_{0 +}^2} - \frac{a^2 \sin^2{(\theta)} \, K \, (L + K)}{\rho_{0 +}^2 \rho_0^2 \, \beta} \, \frac{r + r_+}{r - r_-}\right] \textnormal{d}K + \left[\frac{\rho_+^2}{\rho_{0 +}^2} + \frac{a^2 \sin^2{(\theta)} \, L \, (L + K)}{\rho_{0 +}^2 \rho_0^2 \, \beta} \, \frac{r + r_+}{r - r_-}\right] \textnormal{d}L \\
& \hspace{1.3cm} - a \sin^2{(\theta)} \, \alpha_+ (L + K) \, \textnormal{d}\tilde{\varphi}_+\Biggr) \\ \\
\boldsymbol{m}' & = - \frac{1}{\sqrt{2} \, \overline{\varsigma}} \left(\frac{\textnormal{i} a \sin{(\theta)}}{\alpha_+ \beta \, \rho^2_{0 +}} \, \frac{r + r_+}{r - r_-} \left[L \, \textnormal{d}K - K \, \textnormal{d}L \right] - \rho^2 \, \textnormal{d}\theta + \textnormal{i} \rho^2_0 \sin{(\theta)} \,  \textnormal{d}\tilde{\varphi}_+\right)\\ \\
\overline{\boldsymbol{m}}{\, '} & = \frac{1}{\sqrt{2} \, \varsigma} \left(\frac{\textnormal{i} a \sin{(\theta)}}{\alpha_+ \beta \, \rho^2_{0 +}} \, \frac{r + r_+}{r - r_-} \left[L \, \textnormal{d}K - K \, \textnormal{d}L\right] + \rho^2 \, \textnormal{d}\theta + \textnormal{i} \rho^2_0 \sin{(\theta)} \, \textnormal{d}\tilde{\varphi}_+\right) \, .
\end{split}
\end{equation}
We compute the spin coefficients by inserting the dual co-tetrad (\ref{NPZ}) into the first Maurer--Cartan equation of structure in the Newman--Penrose formalism (\ref{NPMCE}), giving rise to an algebraic system with solution 
\begin{equation}\label{origsc}
\begin{split}
\kappa' & = \sigma' = \lambda' = \nu' = 0 \, , \,\,\,\,\,\,\,\,\,\, \varrho' = - \frac{L - K}{\sqrt{2 \beta} \, \rho \, \varsigma} \, , \,\,\,\,\,\,\,\,\,\, \mu' = - \frac{L + K}{\sqrt{2 \beta} \, \rho \, \varsigma} \, , \,\,\,\,\,\,\,\,\,\, \tau' = - \frac{\textnormal{i} a \sin{(\theta)}}{\sqrt{2} \, {\rho}^2} \, , \\ \\
\epsilon' & = \frac{\alpha_+ (L - K)}{2 \sqrt{2}} \, \partial_r \Omega \, , \,\,\,\,\,\,\,\, \pi' = \frac{\textnormal{i} a \sin{(\theta)}}{\sqrt{2} \, \varsigma^2} \, , \,\,\,\,\,\,\,\,\,\, \gamma' = \frac{\alpha_+ (L + K)}{2 \sqrt{2}} \, \biggl(\frac{2 \textnormal{i} a \cos{(\theta)} \, \Omega}{\rho^2} + \partial_r \Omega\biggr) \, ,\\ \\
\alpha' & = \frac{1}{2 \sqrt{2} \, \varsigma} \Biggl(\textnormal{i} a \sin{(\theta)} \, \biggl[\alpha_+ + \frac{2 r}{\rho^2}\biggr] + \frac{\rho_0^2 \cot{(\theta)}}{\rho^2}\Biggr) \, , \,\,\,\,\,\,\,\,\,\, \beta' = - \frac{1}{2 \sqrt{2} \, \overline{\varsigma}} \biggl(\textnormal{i} a \, \alpha_+ \sin{(\theta)} + \frac{\rho_0^2 \cot{(\theta)}}{\rho^2}\biggr) \, .
\end{split}
\end{equation}

\section{Isolated Horizon Conditions} \label{AD}

\noindent We impose the isolated horizon conditions on the regular Kinnersley tetrad (\ref{NPBV}) by requiring, on the one hand, that the real-valued null vector $\boldsymbol{l'}$, evaluated at $r = r_+$, constitutes the equivalence class of expansion-free null normals (\ref{ECNN}) and, in addition, that it coincides with the generator of the black hole's Killing horizon. On the other hand, we want the associated complex null vectors $\boldsymbol{m'}$ and $\overline{\boldsymbol{m}}{\, '}$ to be tangential to the horizon $2$-spheres, spanning their intrinsic geometry \cite{AFK, ABL}. The first condition on $\boldsymbol{l}'$ is stated as follows. Let $\boldsymbol{\nu}$ be a null normal to $\Delta$, i.e.,
\begin{equation}\label{NCOND}
\left\langle \boldsymbol{\nu}, \boldsymbol{V} \right\rangle = 0 \,\,\,\, \forall \,\,\,\, \boldsymbol{V} \in T\Delta \,\,\,\,\,\,\, \textnormal{and} \,\,\,\,\,\,
\left\langle \boldsymbol{\nu}, \boldsymbol{\nu} \right\rangle = 0 \, ,
\end{equation}
where $\left\langle \cdot \,, \cdot \right\rangle := \boldsymbol{g}\left(\cdot \,, \cdot\right) : \, T\mathfrak{M} \times T\mathfrak{M} \rightarrow \mathbb{R}$ is the canonical scalar product with respect to the metric (\ref{KerrmetrKSC}). From (\ref{NCOND}), we find that the null normal takes the form
\begin{equation*}
\boldsymbol{\nu}_{\, | \, r = r_+} = {\nu^t}_{| \, r = r_+} \alpha_+ L \left(\partial_K + \partial_L\right) 
\end{equation*}
with $\nu^t$ being the time-like component in Boyer--Lindquist coordinates. Evaluating the vector $\boldsymbol{l}'$ given in (\ref{NPBV}) at $r = r_+$, we can immediately show that it is already of this form and, therefore, normal to the horizon. Secondly, we adjust the local isolated horizon surface gravity to the surface gravity of a Kerr black hole, which is defined at space-like infinity, by requiring that the null normal ${\boldsymbol{l}'}_{| \, r = r_+}$ corresponds to the generator of the black hole's Killing horizon 
\begin{equation*}\label{kill}
\boldsymbol{\chi} := \partial_t + \frac{a}{\rho^2_{0 +}} \, \partial_{\varphi} = \alpha_ + L \left(\partial_K + \partial_L\right) \, .
\end{equation*}
We impose this condition by using a class III local Lorentz transformation (\ref{LLTT13})
\begin{equation*}
{\boldsymbol{l}''}_{| \, r = r_+} = \xi \, {\boldsymbol{l}'}_{| \, r = r_+} = \boldsymbol{\chi} 
\end{equation*}
with transformation parameters 
\begin{equation*}
\xi = \sqrt{\frac{2}{\beta_+}} \, \frac{\rho_+ L}{\rho^2_{0 +}} \,\,\,\,\,\, \textnormal{and} \,\,\,\,\,\, \psi = 0 \, .
\end{equation*}
Applying this transformation to (\ref{NPBV}) yields
\begin{equation*} 
\begin{split}
{\boldsymbol{l}''}_{| \, r = r_+} & = \alpha_+ L \left(\partial_K + \partial_L\right) \\ \\
{\boldsymbol{n}''}_{| \, r = r_+} & = \frac{\rho_{0 +}^4}{2 \alpha_+ L \rho_+^4 \Omega_+^2} \left(\partial_K - \partial_L\right) - \frac{a r_+}{\alpha_+ \rho_{0 +}^2 \rho_+^2} \, \partial_{\tilde{\varphi}_+} \\ \\
{\boldsymbol{m}''}_{| \, r = r_+} & = \frac{1}{\sqrt{2} \, \overline{\varsigma}_+} \biggl(\textnormal{i} a \sin{(\theta)} \, \alpha_+ L \left[\partial_K + \partial_L\right] - \partial_{\theta} + \frac{\textnormal{i} \rho_+^2}{\rho^2_{0 +} \sin{(\theta)}} \, \partial_{\tilde{\varphi}_+}\biggr) \\ \\
\overline{\boldsymbol{m}}{\, ''}_{| \, r = r_+} & = - \frac{1}{\sqrt{2} \, \varsigma_+} \biggl(\textnormal{i} a \sin{(\theta)} \, \alpha_+ L \left[\partial_K + \partial_L\right] + \partial_{\theta} + \frac{\textnormal{i} \rho_+^2}{\rho^2_{0 +} \sin{(\theta)}} \, \partial_{\tilde{\varphi}_+}\biggr) \, .
\end{split}
\end{equation*}
Next, we formulate the conditions for the complex pair $(\boldsymbol{m}'', \overline{\boldsymbol{m}}\,'')_{| \, r = r_+}$. Let $\boldsymbol{W}$ be a vector field that is tangential to the horizon $2$-spheres, that is,  
\begin{equation*}
\boldsymbol{W} = W^{\theta} \partial_{\theta} + W^{\tilde{\varphi}_+} \partial_{\tilde{\varphi}_+} \, .
\end{equation*}
In order to bring ${\boldsymbol{m}''}_{| \, r = r_+}$ and ${\overline{\boldsymbol{m}}\,''}_{| \, r = r_+}$ into this form, we use a class I local Lorentz transformation (\ref{LLTT13}) that eliminates the $\partial_{K/L}$-components 
\begin{equation*}
\begin{split}
{\boldsymbol{m}'''}_{| \, r = r_+} & = {\boldsymbol{m}''}_{| \, r = r_+} + \textsf{a} \, {\boldsymbol{l}''}_{| \, r = r_+} = {m'''^{\, \theta}}_{| \, r = r_+} \partial_{\theta} + {m'''^{\, \tilde{\varphi}_+}}_{| \, r = r_+} \partial_{\tilde{\varphi}_+} \\ \\
{\overline{\boldsymbol{m}}\,'''}_{| \, r = r_+} & = {\overline{\boldsymbol{m}}\,''}_{| \, r = r_+} + \overline{\textsf{a}} \, {\boldsymbol{l}''}_{| \, r = r_+} = {\overline{m}\,'''^{\, \theta}}_{| \, r = r_+} \partial_{\theta} + {\overline{m}\,'''^{\, \tilde{\varphi}_+}}_{| \, r = r_+} \partial_{\tilde{\varphi}_+} \, ,
\end{split}
\end{equation*}
where the complex parameter $\textsf{a}$ reads
\begin{equation*}
\textsf{a} = - \frac{\textnormal{i} a \sin{(\theta)}}{\sqrt{2} \, \overline{\varsigma}_+} \, .
\end{equation*}
With this additional transformation, we obtain a Newman--Penrose frame that is adapted to the boundary conditions of Kerr isolated horizons
\begin{equation}\label{NPB3}
\begin{split}
{\boldsymbol{l}'''}_{| \, r = r_+} & = \alpha_+ L \left(\partial_K + \partial_L\right) \\ \\
{\boldsymbol{n}'''}_{| \, r = r_+} & = \frac{\rho^4_{0 +}}{2 \alpha_+ L \rho_+^4 \Omega_+^2} \, \Biggl(\left[1 - \frac{a^2 \sin^2{(\theta)} L^2}{\beta_+ \, \rho_{0 +}^4}\right] \partial_K - \left[1 + \frac{a^2 \sin^2{(\theta)} L^2}{\beta_+ \, \rho_{0 +}^4}\right] \partial_L\Biggr) - \frac{a}{\rho^2_{0 +}} \, \left(1 + \frac{r_+}{\alpha_+ \rho_+^2}\right) \partial_{\tilde{\varphi}_+} \\ \\
{\boldsymbol{m}'''}_{| \, r = r_+} & = \frac{1}{\sqrt{2} \, \overline{\varsigma}_+} \left(- \partial_{\theta} + \frac{\textnormal{i} \rho_{+}^2}{\rho_{0 +}^2 \sin{(\theta)}} \, \partial_{\tilde{\varphi}_+}\right) \\ \\
{\overline{\boldsymbol{m}}\,'''}_{| \, r = r_+} & = - \frac{1}{\sqrt{2} \, \varsigma_+} \left(\partial_{\theta} + \frac{\textnormal{i} \rho_{+}^2}{\rho_{0 +}^2 \sin{(\theta)}} \, \partial_{\tilde{\varphi}_+}\right) \, .
\end{split}
\end{equation} 
Its dual becomes
\begin{equation} \label{FINNPT}
\begin{split}
{\boldsymbol{l}'''}_{| \, r = r_+} & = \frac{\alpha_+ L \, \rho_+^4 \Omega_+^2}{\rho_{0 +}^4} \left(\textnormal{d}K - \textnormal{d}L\right) \\ \\
{\boldsymbol{n}'''}_{| \, r = r_+} & = \frac{1}{2 \alpha_+ L} \, \Biggl(\left[1 + \frac{a^2 \sin^2{(\theta)} \, L^2}{\beta_+ \, \rho_{0 +}^4}\right] \textnormal{d}K + \left[1 - \frac{a^2 \sin^2{(\theta)} \, L^2}{\beta_+ \, \rho_{0 +}^4}\right] \textnormal{d}L\Biggr) \\ \\
{\boldsymbol{m}'''}_{| \, r = r_+} & = \frac{1}{\sqrt{2} \, \overline{\varsigma}_+} \, \Biggl(\frac{\textnormal{i} a \sin{(\theta)} \, L \rho_+^4 \Omega_+^2}{\rho_{0 +}^4} \left[\alpha_+ + \frac{r_+}{\rho_+^2}\right] \left(\textnormal{d}L - \textnormal{d}K\right) + \rho_+^2 \, \textnormal{d}\theta - \textnormal{i} \rho_{0 +}^2 \sin{(\theta)} \, \textnormal{d}\tilde{\varphi}_+\Biggr)\\ \\
{\overline{\boldsymbol{m}}\,'''}_{| \, r = r_+} & = \frac{1}{\sqrt{2} \, \varsigma_+} \, \Biggl(- \frac{\textnormal{i} a \sin{(\theta)} \, L \rho_+^4 \Omega_+^2}{\rho_{0 +}^4} \left[\alpha_+ + \frac{r_+}{\rho_+^2}\right] \left(\textnormal{d}L - \textnormal{d}K\right) + \rho_+^2 \, \textnormal{d}\theta + \textnormal{i} \rho_{0 +}^2 \sin{(\theta)} \, \textnormal{d}\tilde{\varphi}_+\Biggr) \, .
\end{split}
\end{equation} 
The corresponding spin coefficients are determined by transforming (\ref{origsc}) according to (\ref{LLTSC1}) and (\ref{LLTSC3}). We find 
\begin{equation}\label{DPSCOEF}
\begin{split}
{\kappa'''}_{| \, r = r_+} & = {\sigma'''}_{| \, r = r_+} = {\varrho'''}_{| \, r = r_+} = 0 \, , \,\,\,\,\, {\pi'''}_{| \, r = r_+} = \frac{\textnormal{i} a \sin{(\theta)}}{\sqrt{2} \, \varsigma_+} \left(\alpha_+ + \frac{1}{\varsigma_+}\right) , \,\,\,\,\, {\lambda'''}_{| \, r = r_+} = - \frac{a^2 \sin^2{(\theta)}}{\varsigma^2_+} \left(\frac{\alpha_+}{2} + \frac{1}{\varsigma_+}\right) , \\ \\
{\mu'''}_{| \, r = r_+} & = \frac{1}{\rho^2_+} \left(- r_+ + \frac{a^2 \sin^2{(\theta)} \, \alpha_+}{2}\right) , \,\,\,\,\, {\gamma'''}_{| \, r = r_+} = \frac{1}{2 \rho^2_+} \, \Biggl(- \overline{\varsigma}_+ + \frac{r_+ a^2 \sin^2{(\theta)}}{\rho_+^2} - \frac{\alpha_+}{2} \left[\frac{\beta_+ \rho_{0 +}^4}{L^2} + a^2 \sin^2{(\theta)}\right] + \frac{\rho_{0 +}^2}{4 r_+^2 \alpha_+}\Biggr) \, , \\ \\
{\nu'''}_{| \, r = r_+} & = \frac{\textnormal{i} a \sin{(\theta)}}{\sqrt{2} \, \varsigma_+ \, \rho^2_+} \, \Biggl(- 2 \, \overline{\varsigma}_+ + \frac{\textnormal{i} a^3 \cos{(\theta)} \sin^2{(\theta)}}{\rho_+^2} - \alpha_+ \left[\rho_+^2 + \frac{\beta_+ \rho_{0 +}^4}{2 L^2} + \frac{a^2 \sin^2{(\theta)}}{2}\right] + \frac{\rho_{0 +}^2}{4 r_+^2 \alpha_+}\Biggr) \, , \,\,\,\,\, {\tau'''}_{| \, r = r_+} = - \frac{\textnormal{i} a \sin{(\theta)}}{\sqrt{2} \, {\rho}^2_+} \, , \\ \\
{\epsilon'''}_{| \, r = r_+} & = \frac{\alpha_+}{2} \, , \,\,\,\,\, {\alpha'''}_{| \, r = r_+} = \frac{1}{\sqrt{2} \, \varsigma_+} \, \Biggl(\frac{\cot{(\theta)}}{2} + \textnormal{i} a \sin{(\theta)} \, \biggl[\frac{\alpha_+}{2} + \frac{1}{\varsigma_+}\biggr]\Biggr) \, , \,\,\,\,\,
{\beta'''}_{| \, r = r_+} = - \frac{1}{2 \sqrt{2} \, \overline{\varsigma}_+} \bigl(\cot{(\theta)} + \textnormal{i} a \, \alpha_+ \sin{(\theta)}\bigr) \, .
\end{split}
\end{equation}
It can be easily verified that the expansion of the real-valued Newman--Penrose vector ${\boldsymbol{l}'''}_{| \, r = r_+}$ given in (\ref{NPB3}), which constitutes the equivalence class of null normals (\ref{ECNN}), vanishes on the horizon, as in the Newman--Penrose formalism the associated expansion scalar is expressed in terms of the real part of the spin coefficient $\varrho'''$ \cite{AFK}. Thus, using ${\varrho'''}_{| \, r = r_+}$ as stated in (\ref{DPSCOEF}), we directly see that the expansion vanishes on the horizon
\begin{equation*}
\theta_{(\boldsymbol{l}''')| \, r = r_+} = - 2 \, \textnormal{Re}{({\varrho'''}_{| \, r = r_+})} = 0 \, .
\end{equation*}
Moreover, since 
\begin{equation*}
{\kappa'''}_{| \, r = r_+} = {\sigma'''}_{| \, r = r_+} = \textnormal{Im}{({\varrho'''}_{| \, r = r_+})} = 0 \, ,
\end{equation*}
we know that ${\boldsymbol{l}'''}_{| \, r = r_+}$ is geodesic as well as shear- and twist-free. Finally, we remark that the validity of Eq.(\ref{IHMETRICCOND}) can be shown by direct computation and that in the family of Kerr spacetimes, the only equivalence class (\ref{ECNN}) satisfying the isolated horizon condition (v) is the
one containing constant multiples of ${\boldsymbol{l}'''}_{| \, r = r_+}$ \cite{AFK}.

\section{Spin Connection in the Time Gauge} \label{A7}

\noindent The spin connection for Kerr isolated horizons is computed as follows. Dropping the primes and the restrictions in the next formula for readability, we first translate the spin connection $1$-forms $\omega_{(a) (b)} = \gamma_{(a) (b) (c)} \, \boldsymbol{e}^{(c)}$, where $\gamma_{(a) (b) (c)}$ are the Ricci rotation coefficients, into the Newman--Penrose formalism employing (\ref{rrcspc})
\begin{equation} \label{scNPF}
\begin{split}
\omega_{(0) (1)} & =  - 2 \, \textnormal{Re}{(\gamma)} \, \boldsymbol{l} - 2 \, \textnormal{Re}{(\epsilon)} \, \boldsymbol{n} + (\alpha + \overline{\beta}) \, \boldsymbol{m} + (\overline{\alpha} + \beta) \, \overline{\boldsymbol{m}} \\ 
\omega_{(0) (2)} & = \overline{\omega}_{(0) (3)} = - \tau \, \boldsymbol{l} - \kappa \, \boldsymbol{n} + \varrho \, \boldsymbol{m} + \sigma \, \overline{\boldsymbol{m}} \\ 
\omega_{(1) (2)} & = \overline{\omega}_{(1) (3)} = \overline{\nu} \, \boldsymbol{l} + \overline{\pi} \, \boldsymbol{n} - \overline{\mu} \, \boldsymbol{m} - \overline{\lambda} \, \overline{\boldsymbol{m}} \\ 
\omega_{(2) (3)} & = 2 \, \textnormal{i} \, \textnormal{Im}{(\gamma)} \, \boldsymbol{l} + 2 \, \textnormal{i} \, \textnormal{Im}{(\epsilon)} \, \boldsymbol{n} - (\alpha - \overline{\beta}) \, \boldsymbol{m} + (\overline{\alpha} - \beta) \, \overline{\boldsymbol{m}} \, .
\end{split}
\end{equation}
We note that it is sufficient to compute the pullback of these quantities to the isolated horizon $2$-spheres as the entire horizon can be covered by the action of the Lie derivative along its local null normal vector field. Accordingly, inserting the spin coefficients (\ref{DPSCOEF}) and the dual Kinnersley frame (\ref{FINNPT}) into (\ref{scNPF}), we obtain for the spin connection on the isolated horizon $2$-spheres 
\begin{equation} \label{SCONS2}
\begin{split}
\sdpb{\hspace{0.07cm}\omega}'''_{\hspace{0.04cm} (0) (1)} & = \frac{a \sin{(\theta)}}{\rho_+^2} \, \Biggl(a \cos{(\theta)} \, \textnormal{d}\theta + \rho_{0 +}^2 \sin{(\theta)}  \left[\alpha_+ + \frac{r_+}{\rho_+^2}\right] \textnormal{d}\tilde{\varphi}_+\Biggr) \\ \\
\sdpb{\hspace{0.07cm}\omega}'''_{\hspace{0.04cm} (0) (2)} & = \sdpb{\hspace{0.07cm}\omega}'''_{\hspace{0.04cm} (0) (3)} = 0 \\ \\
\sdpb{\hspace{0.07cm}\omega}'''_{\hspace{0.04cm} (1) (2)} & = \sdpb{\hspace{0.07cm}\overline{\omega}}'''_{\hspace{0.04cm} (1) (3)} = \frac{1}{\sqrt{2} \, \overline{\varsigma}_+ \, \rho_+^2} \, \Biggl(\rho_+^2 \left[r_+ + \frac{a^2 \sin^2{(\theta)}}{\overline{\varsigma}_+}\right] \textnormal{d}\theta + \textnormal{i} \rho_{0 +}^2 \sin{(\theta)} \, \biggl[- r_+ + a^2 \sin^2{(\theta)} \left(\alpha_+ + \frac{1}{\overline{\varsigma}_+}\right)\biggr] \, \textnormal{d}\tilde{\varphi}_+\Biggr) \\ \\ 
\sdpb{\hspace{0.07cm}\omega}'''_{\hspace{0.04cm} (2) (3)} & = \frac{\textnormal{i}}{\rho_+^2} \left(- r_+ a \sin{(\theta)} \, \textnormal{d}\theta + \frac{\rho_{0 +}^4 \cos{(\theta)}}{\rho_+^2} \, \textnormal{d}\tilde{\varphi}_+\right) \, ,
\end{split}
\end{equation}
in which the double arrow underneath denotes the horizon $2$-sphere pullback. Since the Ashtekar and Ashtekar--Barbero connection variables are usually defined on space-like hypersurfaces of foliations of globally hyperbolic spacetimes, we now express the spin connection (\ref{SCONS2}) via an orthonormal frame that is fixed in the time gauge (see, e.g., \cite{DS} and references therein). This frame is related to the Newman--Penrose frame via
\begin{equation} \label{rottr}
\boldsymbol{e}\indices{^{(0)}} = \frac{\boldsymbol{l}''' + \boldsymbol{n}'''}{\sqrt{2}} \, , \,\,\,\,\,\, \boldsymbol{e}\indices{^{(1)}} = \frac{\boldsymbol{l}''' - \boldsymbol{n}'''}{\sqrt{2}} \, , \,\,\,\,\,\, \boldsymbol{e}\indices{^{(2)}} = \frac{\boldsymbol{m}''' + \overline{\boldsymbol{m}}{\, '''}}{\sqrt{2}} \, , \,\,\,\,\,\, \textnormal{and} \,\,\,\,\,\, \boldsymbol{e}\indices{^{(3)}} = \frac{\textnormal{i} \, (\boldsymbol{m}''' - \overline{\boldsymbol{m}}{\, '''})}{\sqrt{2}} \, ,
\end{equation}
and the metric is given by the Minkowski metric with signature $(1, 3)$
\begin{equation} \label{locMinkmetric}
\bigl(\tilde{\eta}_{(a) (b)}\bigr) = \begin{pmatrix} 1 & 0 & 0 & 0 \\ 0 & - 1 & 0 & 0 \\ 0 & 0 & - 1 & 0 \\ 0 & 0 & 0 & - 1 \end{pmatrix} \, .
\end{equation}
Substituting (\ref{NPB3}) into (\ref{rottr}) yields 
\begin{equation}\label{ONBV}
\begin{split}
{\boldsymbol{e}^{(0)}}_{| \, r = r_+} & = \frac{\alpha_+ L}{\sqrt{2}} \, \Biggl(\left[1 + \frac{1}{2 \rho_+^2} \left(\frac{\beta_+ \rho_{0 +}^4}{L^2} - a^2 \sin^2{(\theta)}\right)\right] \partial_K + \left[1 - \frac{1}{2 \rho_+^2} \left(\frac{\beta_+ \rho_{0 +}^4}{L^2} + a^2 \sin^2{(\theta)}\right)\right] \partial_L\Biggr) - \frac{a}{\sqrt{2} \rho_{0 +}^2} \left(1 + \frac{r_+}{\alpha_+ \rho_+^2}\right) \partial_{\tilde{\varphi}_+} \\ \\
{\boldsymbol{e}^{(1)}}_{| \, r = r_+} & = \frac{\alpha_+ L}{\sqrt{2}} \, \Biggl(\left[1 - \frac{1}{2 \rho_+^2} \left(\frac{\beta_+ \rho_{0 +}^4}{L^2} - a^2 \sin^2{(\theta)}\right)\right] \partial_K + \left[1 + \frac{1}{2 \rho_+^2} \left(\frac{\beta_+ \rho_{0 +}^4}{L^2}  + a^2 \sin^2{(\theta)}\right)\right] \partial_L\Biggr) + \frac{a}{\sqrt{2} \rho_{0 +}^2} \left(1 + \frac{r_+}{\alpha_+ \rho_+^2}\right) \partial_{\tilde{\varphi}_+} \\ \\
{\boldsymbol{e}^{(2)}}_{| \, r = r_+} & = - \frac{r_+}{\rho_+^2} \, \partial_{\theta} - \frac{a \cot{(\theta)}}{\rho_{0 +}^2} \, \partial_{\tilde{\varphi}_+} \\ \\
{\boldsymbol{e}^{(3)}}_{| \, r = r_+} & = \frac{a \cos{(\theta)}}{\rho_+^2} \, \partial_{\theta} - \frac{r_+}{\rho_{0 +}^2 \sin{(\theta)}} \, \partial_{\tilde{\varphi}_+} \, .
\end{split}
\end{equation}
As we work only on the horizon $2$-spheres, the orthonormal frame is in the time gauge when $\bigl\langle {\boldsymbol{e}^{(0)}}_{| \, r = r_+}, \boldsymbol{W} \bigr\rangle = 0$ for all $\boldsymbol{W} = W^{\theta} \partial_{\theta} + W^{\tilde{\varphi}_+} \partial_{\tilde{\varphi}_+} \in TS^2$. This condition leads to a time-like basis vector of the general form
\begin{equation*}
{\boldsymbol{e}^{(0)}}_{| \, r = r_+} = {e^{(0) K}}_{| \, r = r_+} \partial_K + {e^{(0) L}}_{| \, r = r_+} \partial_L + \frac{a L \Omega_+^2 \rho_+^4}{\rho_{0 +}^6} \left(\alpha_+ + \frac{r_+}{\rho_+^2}\right) \Bigl[{e^{(0) L}}_{| \, r = r_+} - {e^{(0) K}}_{| \, r = r_+}\Bigr] \, \partial_{\tilde{\varphi}_+} \, .
\end{equation*}
Direct computation shows that the time-like basis vector ${\boldsymbol{e}^{(0)}}_{| \, r = r_+}$ of the orthonormal tetrad (\ref{ONBV}) is already of this form.
Transforming the spin connection (\ref{SCONS2}) into the orthonormal frame according to the transformation law for connections 
\begin{equation*}
\sdpb{\hspace{0.07cm}{\omega}_{\, (a) (b)}} = \tilde{\eta}\indices{_{(a) (c)}} \,  \tilde{\eta}\indices{_{(h) (b)}} \, \eta\indices{^{(d) (e)}} \, \Bigl[\Lambda\indices{^{(c)}_{(d)}} \, \sdpb{\hspace{0.07cm}{\omega}'''_{\hspace{0.04cm} (e) (f)}} \,  \eta\indices{^{(f) (g)}} \, \bigl(\Lambda\indices{^{(h)}_{(g)}}\bigr)^T - \bigl(\textnormal{d}\Lambda\indices{^{(c)}_{(d)}}\bigr) \, \bigl(\Lambda\indices{^{(h)}_{(e)}}\bigr)^T \, \Bigr] \, ,
\end{equation*}
where the transformation matrix is given by (see (\ref{rottr}) and Appendix \ref{AA})
\begin{equation*} 
\Bigl(\Lambda\indices{^{(a)}_{(b)}}\Bigr) = \frac{1}{\sqrt{2}} \, \begin{pmatrix} 1 & 1 & 0 & 0 \\ - 1 & 1 & 0 & 0 \\ 0 & 0 & - 1 & - 1 \\ 0 & 0 & \textnormal{i} & - \textnormal{i} \end{pmatrix} 
\end{equation*}
and the metrics $\boldsymbol{\eta}$ and $\tilde{\boldsymbol{\eta}}$ by (\ref{locNPmetric}) and (\ref{locMinkmetric}), respectively, results in the expressions (\ref{pbsc}).

\section{Ashtekar and Ashtekar--Barbero Formulations} \label{AE}

\noindent In the Ashtekar formulation, instead of using a tetrad frame $(\boldsymbol{e}^{(a)})$ and six $\mathfrak{so}_{1, 3}(\mathbb{R})$-valued spin connection $1$-forms $\omega^{(a) (b)}$, with $a, b \in \{0, 1, 2, 3\}$, one employs a triad $(\boldsymbol{e}^{(i)})$ and three self-dual $\mathfrak{sl}_{2}(\mathbb{C})$-valued connection $1$-forms $A^{(i)}_+$, $i \in \{1, 2, 3\}$, as configuration variables (for a detailed review see, e.g., \cite{Rov1}). The self-dual connection, commonly known as Ashtekar connection, is defined by \cite{A}
\begin{equation}\label{selfdualAsh}
A_{+}^{(i)} := P_{+ \,\,\,\,\,\, (a)(b)}^{(i)} \, \omega\indices{^{(a) (b)}} \, , 
\end{equation}
where
\begin{equation}\label{sdacproj}
P_{+ \,\,\,\,\,\, (j)(k)}^{(i)} := \frac{1}{2} \, \epsilon\indices{^{(i)}_{(j)(k)}} \, , \,\,\,\,\,\,\,\, P_{+ \,\,\,\,\,\, (0)(k)}^{(i)} = - P_{+ \,\,\,\,\,\, (k)(0)}^{(i)} := \frac{\textnormal{i}}{2} \, \delta^{(i)}_{(k)}
\end{equation}
is a projection homomorphism reading out the self-dual part of the spin connection. Note that, since the $\mathfrak{so}_{1, 3}(\mathbb{R})$ Lie algebra cannot be decomposed into a direct sum of self-dual and anti-self-dual algebras, one has to work with the complexified Lie algebra $\mathfrak{so}_{1, 3}(\mathbb{C})$ for which such a decomposition exists, namely
\begin{equation*}
\mathfrak{so}_{1, 3}(\mathbb{C}) = \mathfrak{so}_{1, 3}(\mathbb{C})_+ \oplus \mathfrak{so}_{1, 3}(\mathbb{C})_- 
\end{equation*}
with the self-dual and anti-self-dual algebras
\begin{equation*}
\mathfrak{so}_{1, 3}(\mathbb{C})_+ := \{\boldsymbol{\tau} \in \mathfrak{so}_{1, 3}(\mathbb{C}) \,\, | \star\boldsymbol{\tau} = \textnormal{i} \boldsymbol{\tau} \} \,\,\,\,\,\, \textnormal{and} \,\,\,\,\,\, \mathfrak{so}_{1, 3}(\mathbb{C})_- := \{\boldsymbol{\tau} \in \mathfrak{so}_{1, 3}(\mathbb{C}) \,\, | \star\boldsymbol{\tau} = - \textnormal{i} \boldsymbol{\tau} \} \, , 
\end{equation*}
where $\boldsymbol{\tau}$ is an eigenfunction of the $\star$-operator. Then, the spin connection can be split into a self-dual and an anti-self-dual part 
\begin{equation*}
\boldsymbol{\omega} = \boldsymbol{\omega}_+ + \boldsymbol{\omega}_- \, ,
\end{equation*}
in which both $\boldsymbol{\omega}_+$ and $\boldsymbol{\omega}_-$ contain the same information as $\boldsymbol{\omega}$ itself. The Ashtekar connection (\ref{selfdualAsh}) can be generalized to an $\mathfrak{su}_{2}(\mathbb{C})$-valued $1$-form, the so-called Ashtekar--Barbero connection $A^{(i)}_{\gamma}$, replacing the imaginary unit in the projector (\ref{sdacproj}) by a complex-valued parameter $\gamma$, known as Barbero--Immirzi parameter. Different values for $\gamma$ yield equivalent classical theories, however, their quantum theories are unitarily inequivalent, resulting in ambiguous physical predictions. In the framework of loop quantum gravity, one usually restricts $\gamma$ to values in $\mathbb{R} \backslash \{0\}$ \cite{Barb}, which in turn leads to real-valued SU($2$) connections. This restriction arises from the current level of development in functional analysis, where mathematical methods are worked out in detail only for real-valued connections. 

\end{appendix}

\end{document}